\documentclass[preprint]{aastex}
\graphicspath{{../Pics/}}

\usepackage{graphicx}
\usepackage{dcolumn}
\usepackage{bm}
\usepackage{color}
\usepackage[T1]{fontenc}

\definecolor{myorange}{rgb}{1,0.6,0}
\definecolor{myblue}{rgb}{0.1,0,0.7}
\definecolor{myred}{rgb}{0.8,0,0.2}
\definecolor{mygreen}{rgb}{0.2,0.7,0.55}

\shorttitle{Perpendicular shocks in mixed plasmas}
\shortauthors{Stockem et al.}

\begin{document}

\title{Acceleration in perpendicular relativistic shocks for plasmas consisting of leptons and hadrons}

\author{A. Stockem\altaffilmark{1}, F. Fi\'uza\altaffilmark{1}, R.A. Fonseca\altaffilmark{1,2} and L.O. Silva\altaffilmark{1}}
\affil{$^1$GoLP/Instituto de Plasmas e Fus\~{a}o Nuclear - Laborat\'{o}rio Associado, Instituto Superior T\'{e}cnico, Lisboa, Portugal}
\affil{$^2$DCTI, ISCTE - Lisbon University Institute, Portugal}
\email{anne.stockem@ist.utl.pt}

\begin{abstract}
We investigate the acceleration of light particles in perpendicular shocks for plasmas consisting of a mixture of leptonic and hadronic particles. Starting from the full set of conservation equations for the mixed plasma constituents, we generalize the magneto-hydrodynamical jump conditions for a multi-component plasma, including information about the specific adiabatic constants for the different species. The impact of deviations from the standard model of an ideal gas is compared in theory and particle-in-cell simulations, showing that the standard-MHD model is a good approximation. The simulations of shocks in electron-positron-ion plasmas  are for the first time multi-dimensional, transverse effects are small in this configuration and 1D simulations are a good representation if the initial magnetization is chosen high. 1D runs with a mass ratio of 1836 are performed, which identify the Larmor frequency \(\omega_{ci}\) as the dominant frequency that determines the shock physics in mixed component plasmas. The maximum energy in the non-thermal tail of the particle spectra evolves in time according to a power-law \(\propto t^\alpha\) with \(\alpha\) in the range \(1/3 < \alpha < 1\), depending on the initial parameters. A connection is made with transport theoretical models by \cite{D83} and \cite{GS11}, which predict an acceleration time \(\propto \gamma\) and the theory for small wavelength scattering by \cite{KR10}, which predicts a behavior rather as \(\propto \gamma^2\). Furthermore, we compare different magnetic field orientations with \(\mathbf B_0\) inside and out of the plane, observing qualitatively different particle spectra than in pure electron-ion shocks.

\end{abstract}


\maketitle

\section{Introduction}

Shock acceleration has received considerable attention in recent years, due to the possibility of accelerating charged particles to very high energies. The investigation of shocks in pair plasmas has been motivated by \cite{A83}, arguing that these plasma constituents are dominant in some astrophysical scenarios. These scenarios are also convenient for numerical simulations (e.\,g.\ \cite{CS08}, \cite{S08b}, \cite{NN09}), due to the fact that numerical simulations including heavier particles are more demanding due to the disparity of typical length and time scales. Simulations of electron-ion shocks are mostly performed with a reduced mass ratio (e.\,g.\ \cite{HS02}, \cite{HH04}, \cite{S08a}, \cite{MF09}, \cite{KT10}). Recently, \cite{H11} studied the full development and relaxation process of an electron-ion shock in a three-dimensional simulation, using a mass ratio \(m_i/m_e = 16\), in a two-dimensional spatial configuration, \cite{SS11} studied electron-ion shocks for the first time with mass ratios \(m_i / m_e = 1000\).

\cite{S08a} showed that electron-ion shocks behave similarly as electron-positron shocks on large time scales, because the particle rest mass becomes negligible in comparison with the relativistic mass once the stage of full downstream thermalization has been obtained. This is supported by the fact that both particle components show comparable energy spectra \cite[]{MF09}, and facilitates the comparison with theoretical models, as the standard jump conditions for a single-fluid, derived by \cite{BM76}, can be applied.
But this is true only for initially unmagnetized or quasi-parallel shocks. The investigation of strongly magnetized perpendicular shocks, that we perform in this paper, shows a different picture. The compression ratio is significantly increased in the presence of a heavy particle component and the shock front propagates at a lower velocity. The results are in good agreement with our analytical derivation of the jump conditions for perpendicular shocks in a multiple species plasma. For this analysis, the only assumption we make is that the downstream density profile is similar for all particle components, which happens after a few \(\omega_{pe}^{-1}\), even if the ions have not thermalized yet.

There still exists a gap between analytical acceleration models and numerical simulations or observation data of perpendicular shocks, as very large amplitudes for the magnetic turbulence are needed to enable multiple scatterings of the particles in the shock and to form the observed energy spectra \cite[]{A07}. Since the acceleration process cannot be explained by a simple model, \cite{A07} suggested a mixture of diffusive Fermi acceleration and an additional acceleration process, where heavy ions play a major role. The latter process provides a mechanism to produce a broad energy range in plasmas, where pairs dominate by number and where ions are energetically dominant, explaining the observed range of optical, X- and \(\gamma\)-radiation in Pulsar Wind Nebulae, but still the spectra are not all in agreement with observations. A characteristic of pure electron-positron perpendicular shocks is, that no evidence has been found for the existence of a non-thermal population (e.\,g.\ \cite{LA88}, \cite{GH92}). This population appears only if the initial magnetic field has an oblique structure \cite[]{SS09} or in the presence of an ion population (\cite{HG92}, \cite{AA06}). In the latter case, \cite{HA91} found that the light plasma species gains energy from the heavy species due to the synchrotron maser instability. The gyrating ions emit magnetosonic waves, which are absorbed preferentially by positrons, accelerating them to non-thermal energies. A ratio \(n_i m_i / n_e m_e > 10\) is necessary to achieve efficient acceleration, as demonstrated by \cite{AA06} in a 1D simulation with \(m_i/m_e = 100\). The left-handed orientation of the waves, facilitates the energization of the positrons, which is why the electron spectrum in such a configuration was not observed to reach the same level as that of the positrons. \cite{HG92} suggested that a realistic mass ratio \(m_i/m_e = 1836\), will have the same accelerating effect on the electrons. We confirm that the electron tail is stronger in this case. However, we observe the acceleration efficiency to be not only a function of the ion mass but also the total magnetization.

The temporal evolution of the maximum energy is investigated for different ion to electron density ratios and it is found to be consistent with the acceleration model due to multiple scattering in small wavelength turbulence \cite[]{KR10}.

This paper is structured as follows. The physical scenario is described in Section \ref{sec1} and the jump conditions are presented for a perpendicular shock in a plasma consisting of mixed particle constituents with different energy spectra. In Section \ref{sec3} the simulation results are compared with the theory from the previous section. We discuss first the differences in the particle spectra and their effects on the jump conditions, where we vary the initial ion kinetic energy ratio for a constant magnetization, which leads to the same jump conditions in the standard MHD model. After, we discuss the advanced model for a wide range of parameters with decreasing magnetization. Finally, the effect of the magnetic field orientation is discussed briefly and the main results of the simulations are discussed and summarized in Section \ref{sec:dis}.

\section{Theory for mixed particle components}\label{sec1}
\begin{figure}[ht!]
\begin{center}
\includegraphics[width=11cm]{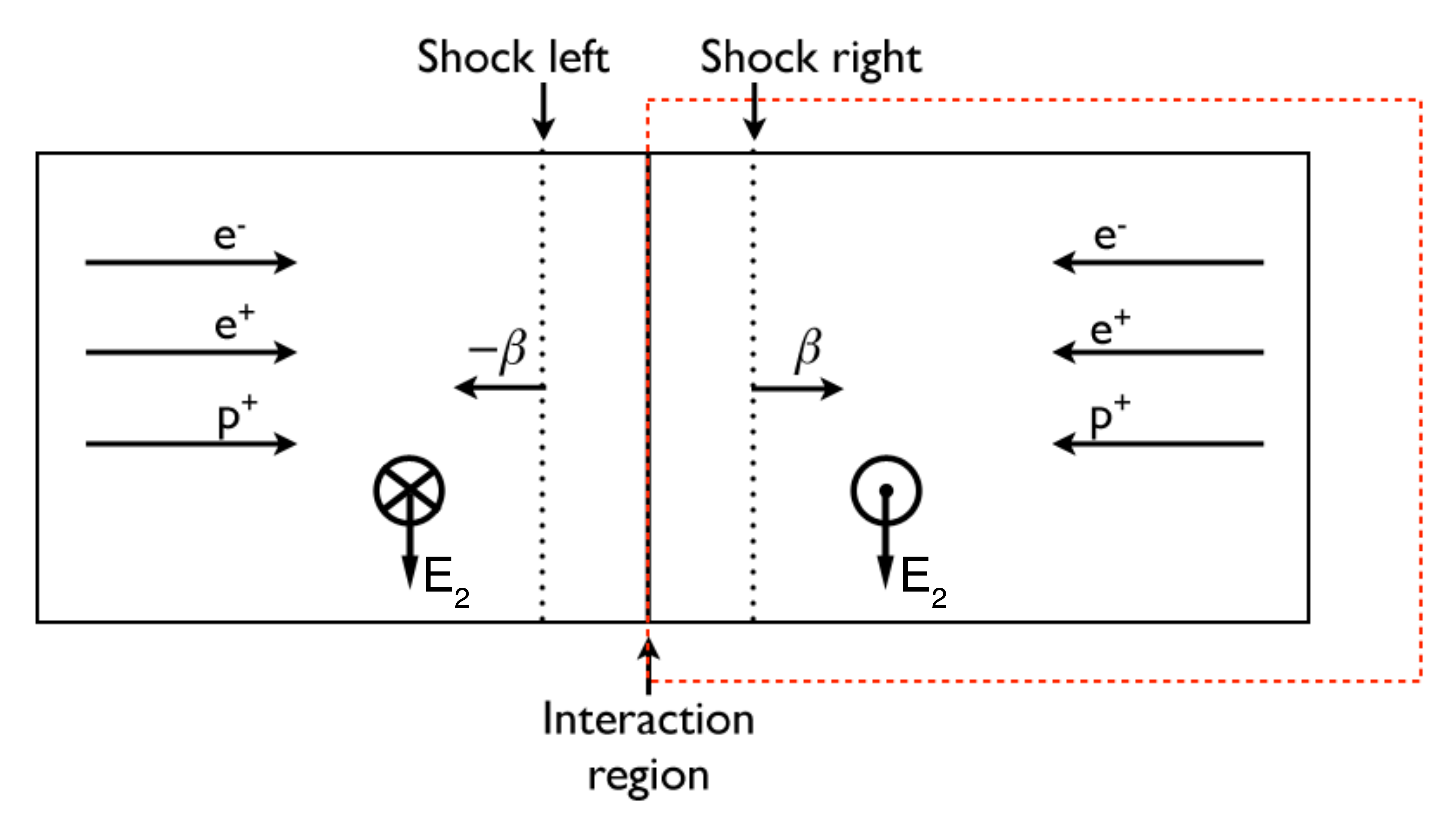}
\end{center}
\vspace{-12pt}
\caption{Shock setup and definition of the right-hand-side (red box).}\label{fig:explain}
\end{figure}

We investigate the interaction of two counterstreaming beams, where each stream is charge-neutral and consisting of a mixture of electrons, positrons and ions, in a constant perpendicular magnetic field. This leads to the formation of two shocks, each propagating in the opposite direction of the incoming upstream beam, away from the interaction region. The following analytical model describes the quasi-steady state after the shock is formed and our discussions throughout this paper are limited to the description of the shock propagating to the right-hand-side (see Figure \ref{fig:explain}). We adopt the syntax of \cite{ZK05} denoting quantities measured in their rest frame with a single index \(Q_i\) and quantities measured in the rest frame \(j\) by \(Q_{ij}\) with \(i,\,j\) = 1, 2, s, denoting the upstream, downstream and shock frame, respectively. The additional index separated by a comma \(Q_{i,a}\), \(Q_{ij,a}\) with \(a=e-\), \(e+\), \(p+\) specifies the species, electron, positron, ion, respectively. As the theory is compared with the simulation results in the following section, the calculations are performed in the simulation frame, which coincides with the downstream frame. 
In this frame, the different species can be treated equally with \( \beta_2 =v_2= 0\) and \(\gamma_2 = 1\). Moreover, all three components have the same upstream velocity \(\beta_{12} = v_{12} / c <0\) and Lorentz factor \(\gamma_{12}\) and the shock propagates with \(\beta := \beta_{s2} = v_{s2}/c>0\) and its associated Lorentz factor \(\gamma := \gamma_{s2}\).

\begin{figure}[ht!]
\begin{center}
\includegraphics[width=8cm]{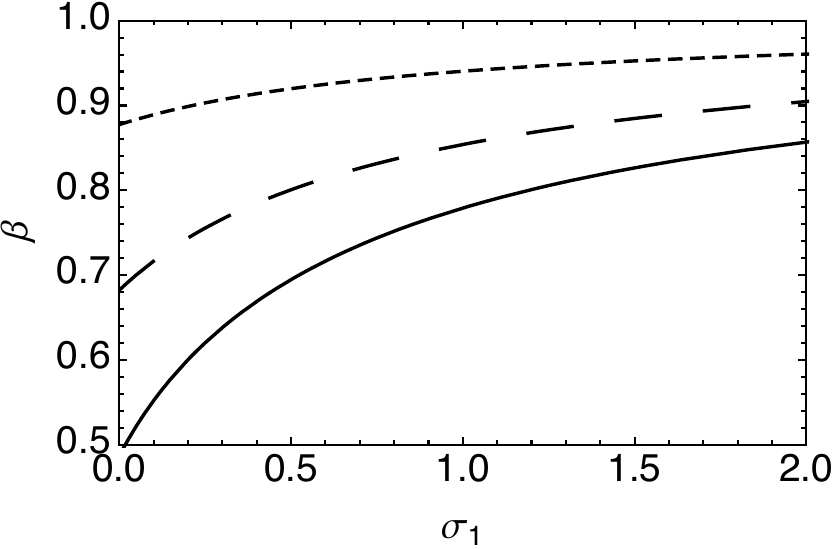}
\includegraphics[width=8cm]{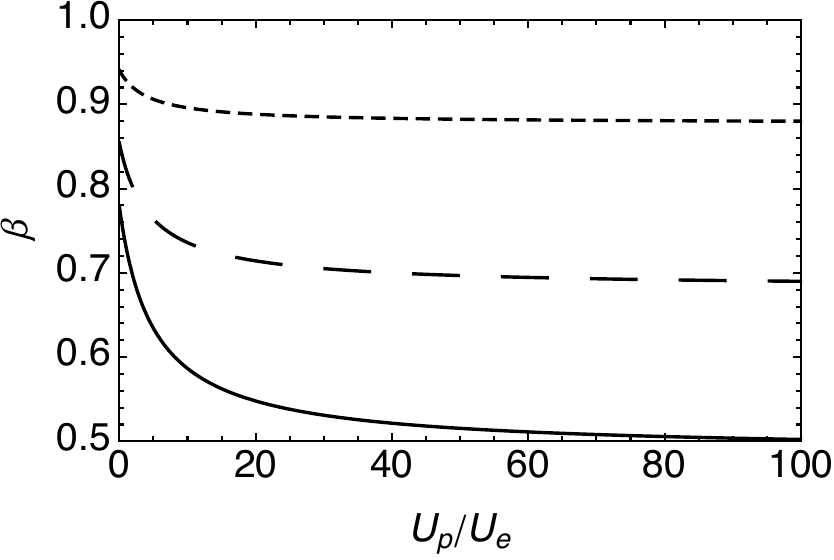}
\end{center}
\vspace{-12pt}
\caption{Shock speed \(\beta\) against (a) the total magnetization \(\sigma_1\) and (b) the fraction of upstream kinetic energy carried by hadrons \(U_p/U_e = m_p n_{12,p+} / (m_e n_{12,e-})\) with \(m_p/m_e=100\) and \(\sigma_e = 2\) for \(\Gamma_{e-} = 1.5\) (solid), 1.7 (large dashed), 1.9 (small dashed), assuming equal particle spectra.}\label{fig:theoretical}
\end{figure}

Because of the charge-neutrality condition, the initial upstream densities of positrons and ions sum up to the electron density \(n_{12}:=n_{12,e-} = n_{12,e+}  + n_{12,p+} \). The simulations show that the downstream densities of the three species behave similarly after a few \(\omega_{pe}^{-1}\), even if the ions have not thermalized yet, therefore we assume  \(n_{2,e-} /  n_{12,e-}  = n_{2,e+}/ n_{12,e+} + n_{2,p+}  / n_{12,p+}\). With these assumptions the jump conditions can be derived in a similar way as for a single fluid.
The detailed derivation of the jump conditions is presented in the Appendix. For the remainder of the paper, we will use the shock speed, as a function of the key parameters, given by
\begin{eqnarray}\label{exact}
 && \gamma_{12}  \left[ 1+ \beta | \beta_{12} | - \frac{ |\beta_{12} |}{\beta \gamma^2} \sigma_1 \right] \left[ 1+  \frac{w_3}{w_4} \frac{\Gamma_{e-}}{\Gamma_{e-}-1} \beta^2 \gamma^2  \right] - \frac{w_2}{w_1} \nonumber \\
 &&  \qquad- \frac{w_3}{w_4} \frac{\Gamma_{e-}}{\Gamma_{e-}-1} \frac{ \beta \gamma_{12}}{ |\beta_{12} | + \beta} \left[ ( |\beta_{12}  |+ \beta)^2 \gamma^2 - \frac{\sigma_1  |\beta_{12} |}{2\beta^2} \left\{  |\beta_{12} | (1+ \beta^2) + 2 \beta \right\} \right] = 0
\end{eqnarray}
and the approximation of Equation (\ref{exact}) for highly relativistic upstream Lorentz factors \(\gamma_{12} \gg 1\)
\begin{equation}\label{bineq}
	\beta^2 (1+\sigma_1) \left(1- \frac{w_3}{w_4} \frac{\Gamma_{e-}}{\Gamma_{e-}-1} \right) + \beta \left[ 1+ \frac{\sigma_1}{2} \frac{w_3}{w_4} \frac{\Gamma_{e-}}{\Gamma_{e-}-1} - \frac{1}{\gamma_{12} } \frac{w_2}{w_1} \right] - \sigma_1 \left(1 - \frac{w_3}{2w_4} \frac{\Gamma_{e-}}{\Gamma_{e-}-1} \right) = 0.
\end{equation}
Equation (\ref{bineq}) reduces to Equation (16) of \cite{GH92} in the limit of equal downstream spectra, where in Equation (\ref{bineq}) it is important to keep in mind that the effective magnetization \(\sigma_1 = \sigma_{tot}\), defined in Equation (\ref{def:sigmatot}), is considered, containing the contributions of all particle components. The jump conditions are determined by the parameter \(\sigma_1\), and the downstream adiabatic constant \(\Gamma_{e-}\). The dependence of the shock speed, defined by Equation (\ref{bineq}), on the total magnetization and the initial ion to electron kinetic energy fraction \(m_p n_{12,p+} / (m_e n_{12,e-})\) is demonstrated in Figure \ref{fig:theoretical}. For the sake of simplicity equal particle spectra are chosen with an adiabatic constant \(\Gamma_{e-} = 3/2\). The shock speed is increased if the total magnetization is increased and decreased with increasing initial ion kinetic energy ratio.

The jump conditions for the density and magnetic field ratios obtained from Equations (\ref{Jump_shock1}) and (\ref{Jump_shock2}) are equal and given by
\begin{equation}\label{densityjump}
	\frac{n_2}{n_{12}} = \frac{n_{2,e-}}{n_{12,e-}} = 1 + \frac{|\beta_{12}|}{\beta} = \frac{B_2}{B_{12}}.
\end{equation}
%


\subsection{The role of the adiabatic constant in the shock properties}
\begin{figure}[ht!]
\begin{center}
\includegraphics[width=8cm]{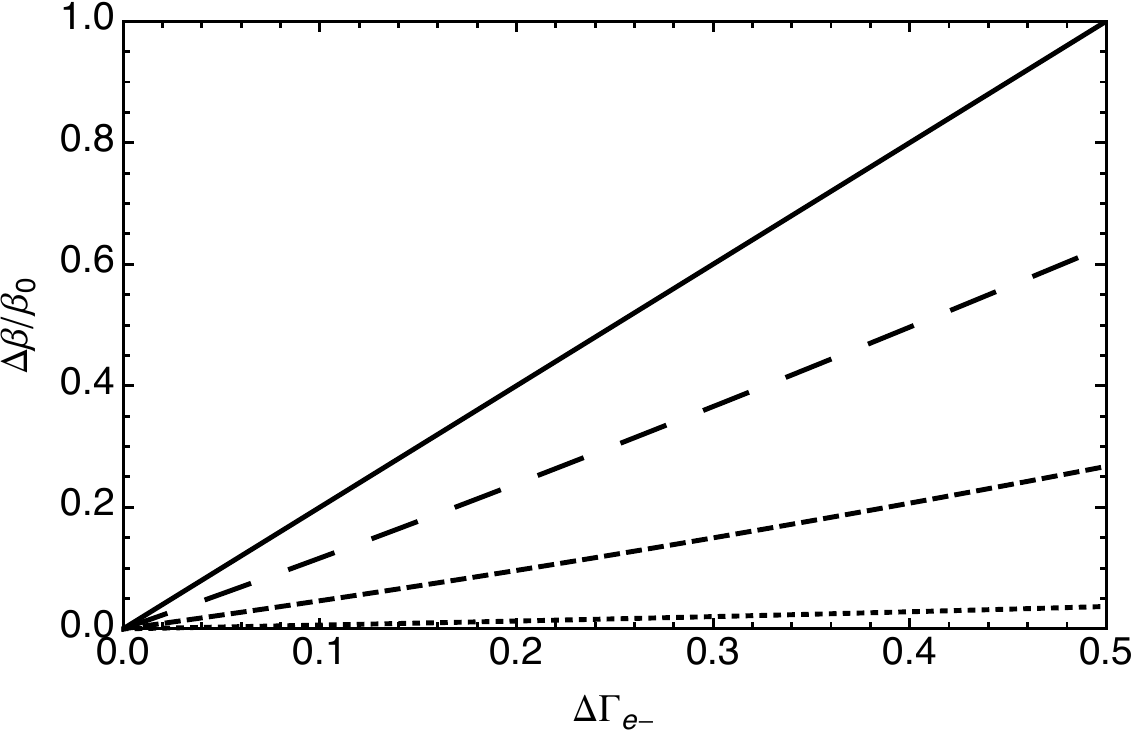}
\end{center}
\vspace{-12pt}
\caption{Deviations in the shock speed \(\Delta \beta = \beta_0 - \beta\), with \(\beta_0 \) the shock speed for  \(\Gamma_{e-} = 3/2 \), according to variations in the adiabatic constant \(\Delta \Gamma_{e-} = \Gamma_{e-} - 3/2\) for magnetizations \(\sigma_1 = 0\) (solid), 0.2 (large dashed), 1 (small dashed), 10 (dotted).}\label{fig:adiabat}
\end{figure}

One of the objectives of this work is to determine the impact of the real particle distributions on the jump conditions. For this, the downstream adiabatic constants and pressure densities have to be determined from the simulation data. In the previous section, the adiabatic constant has been defined for each species as a relation between the energy, pressure and spatial densities, which are defined by
\begin{eqnarray}
	e_{i,a}& := &  2 \pi  \, m_a c^2 \, \int_1^\infty  d\gamma \, \gamma^2 \, f_{i,a}(\gamma) \label{energy} \\
	p_{i,a} & := & \pi  \, m_a c^2  \, \int_1^\infty d\gamma \, (\gamma^2-1) \, f_{i,a}(\gamma) \label{pressure} \\
	n_{i,a} & := & 2 \pi \, \int_1^\infty  d\gamma \, \gamma \, f_{i,a}(\gamma) \label{density}
\end{eqnarray}
in a two-dimensional geometry. After the particle distribution has been determined from the simulation data, the pressure densities and adiabatic constants \(\Gamma_{a} = 1+ p_{2,a} / (e_{2,a} - n_{2,a} m_{a} c^2)\) can be determined.
The distribution functions are found to be fitted well by a Maxwellian for low energies plus a high-energy power-law tail and an exponential cut-off \cite[e.g.][]{S08b}
\begin{equation}\label{fittingfunction}
	f(\gamma) = \gamma^{-1} \frac{\partial n}{\partial \gamma}  = C_1 \left[ \exp \left[-\gamma / \Delta \gamma \right] + C_2 \gamma^{-(1+\alpha)} \, \min \! \left\{ 1, \exp \left[-(\gamma- \gamma_{cut} ) / \Delta \gamma_{cut} \right] \right\} \right]
\end{equation}
with \(C_2 = 0\) for \(\gamma< \gamma_{min}\). An analytical expression of the densities (\ref{energy})-(\ref{density}) is provided in \cite{SF11}. A parameter study of Equation (\ref{bineq}) is presented in Figure \ref{fig:adiabat} showing the variation of the shock speed with the downstream adiabatic constant for a particular initial magnetization. If the magnetic field is strong, the impact of the change in the adiabatic constant, which is determined by the shape of the distribution function, is low, unlike what has been observed for unmagnetized scenarios in \cite{SF11}. In the unmagnetized case, the deviation in the shock speed will be 20\%, whereas it is just 12\% for \(\sigma_1 =0.2\) or 5\% for \(\sigma_1 =1\).

\section{Simulations of highly magnetized perpendicular shocks}\label{sec3}

To study the effect of mixed plasma components on the jump conditions and to test the theory developed in the previous section, we use 1D and 2D particle-in-cell (PIC) simulations, which we perform with the kinetic PIC code OSIRIS \cite[]{FS02,FM08}. We have found that the setup described in Figure \ref{fig:explain} is more appropriate for the numerical study of these shocks, since it avoids boundary condition issues. In this work, we employ a setup reproducing the model of Figure \ref{fig:explain}, using two counterstreaming beams, so that two shocks are formed, propagating in opposite directions, away from the interaction region. The background magnetic field is constant in time and changes its sign according to the sign of the upstream velocities \(\pm \beta_{12}\) of the opposite beams. The motional electric field \(\mathbf E_{12} = - \bm \beta_{12} \times \mathbf B_{12}\) is thus constant over the entire simulation box. Because of the symmetry of the formation of the two shocks, we limit our discussions to the right-hand side of the simulation box (see Figure \ref{fig:explain}).

\subsection{Varying the ion kinetic energy for a constant magnetization}

By adjusting the ion mass and density ratios, different ion kinetic energy ratios \(U_p/U_e = n_{12,p+} m_p / n_{12,e-} m_e\) can lead to the same total magnetization \( \sigma_1 =  \sigma_e / [2 + ( m_p / m_e -1 ) n_{12,p+} / n_{12,e-}] \), with the same jump conditions in the standard MHD model. In this section we discuss the differences in the spectra and their effects on the jump conditions. Due to the constraint \(0 \leq n_{12,p+}/n_{12,e-} \leq 1 \), the difference between the lowest and largest value of the ion kinetic energy ratio for constant \(\sigma_1\) and \(\sigma_e\) is limited to \(\Delta (U_p/U_e) = 1\). The total magnetization is chosen high enough (\(\sigma_1 >10^{-3}\) \cite[]{S05}) to suppress the Weibel instability, so that it can be excluded as the dominant driver for shock formation. A discussion of the role of the Weibel instability in baryon-loaded plasmas is presented in \cite{FS06}.

\begin{figure}[ht!]
\begin{center}
\includegraphics{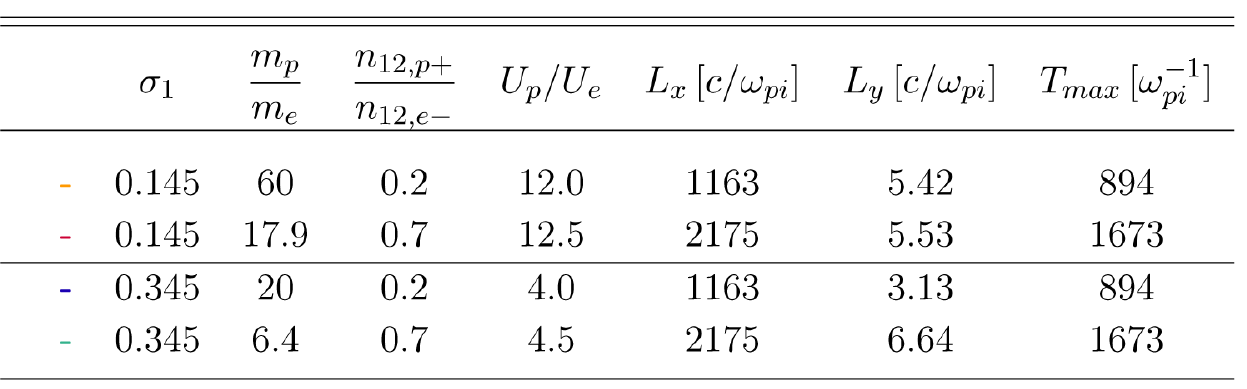}
\end{center}
\vspace{-18pt}
\caption{Simulation parameters}\label{tab:simpar}
\end{figure}

We performed two sets of simulations for magnetizations \(\sigma_1 = 0.345\) with ion kinetic energy ratios \(U_p / U_e = 4.0\) and 4.5 and   \(\sigma_1 = 0.145\) with \(U_p / U_e = 12.0\) and 12.5. The details are given in Fig. \ref{tab:simpar}. Global parameters are the time step \(\Delta t = 0.2214 \, \omega_{pe}^{-1} =  3.9 \, \textrm{ps} \times \sqrt{n_{12,e-} [\textrm{m}^{-3}]} \) with electron plasma frequency \(\omega_{pe} = \sqrt{4 \pi n_{12,e-} e^2 / m_e} = 5.64 \times 10^7 \, \textrm{s}^{-1} \times \sqrt{n_{12,e-} [\textrm{m}^{-3}]} \), the cell size \(\Delta x = \Delta y = 0.44 \, c / \omega_{pe} = 2.34 \, \textrm{m} / \sqrt{n_{12,e-} [\textrm{m}^{-3}]} \) with \(3\times 3\) particles per cell and species, and the magnetic field amplitude \(|\mathbf B_{12}| = 8.94 \, m_e c \omega_{pe} / e = 2.9 \, \textrm{mT} \times \sqrt{n_{12,e-} [\textrm{m}^{-3}]}  \). The relativistic ion Larmor radius is defined as \(r_{Li}  \approx c/ \omega_{ci} = m_p c^2 \gamma_{12} /(e |\mathbf B_{12}|) = 1.9 \, \textrm{km} \times (m_p/m_e) / \sqrt{n_{12,e-} [\textrm{m}^{-3}]}\). Particles are symmetrically injected from both sides of the two-dimensional simulation box.

\subsubsection{Analysis of the particle spectra}\label{sec:spectra}

\begin{figure}[ht!]
\begin{center}
\includegraphics[width=7cm]{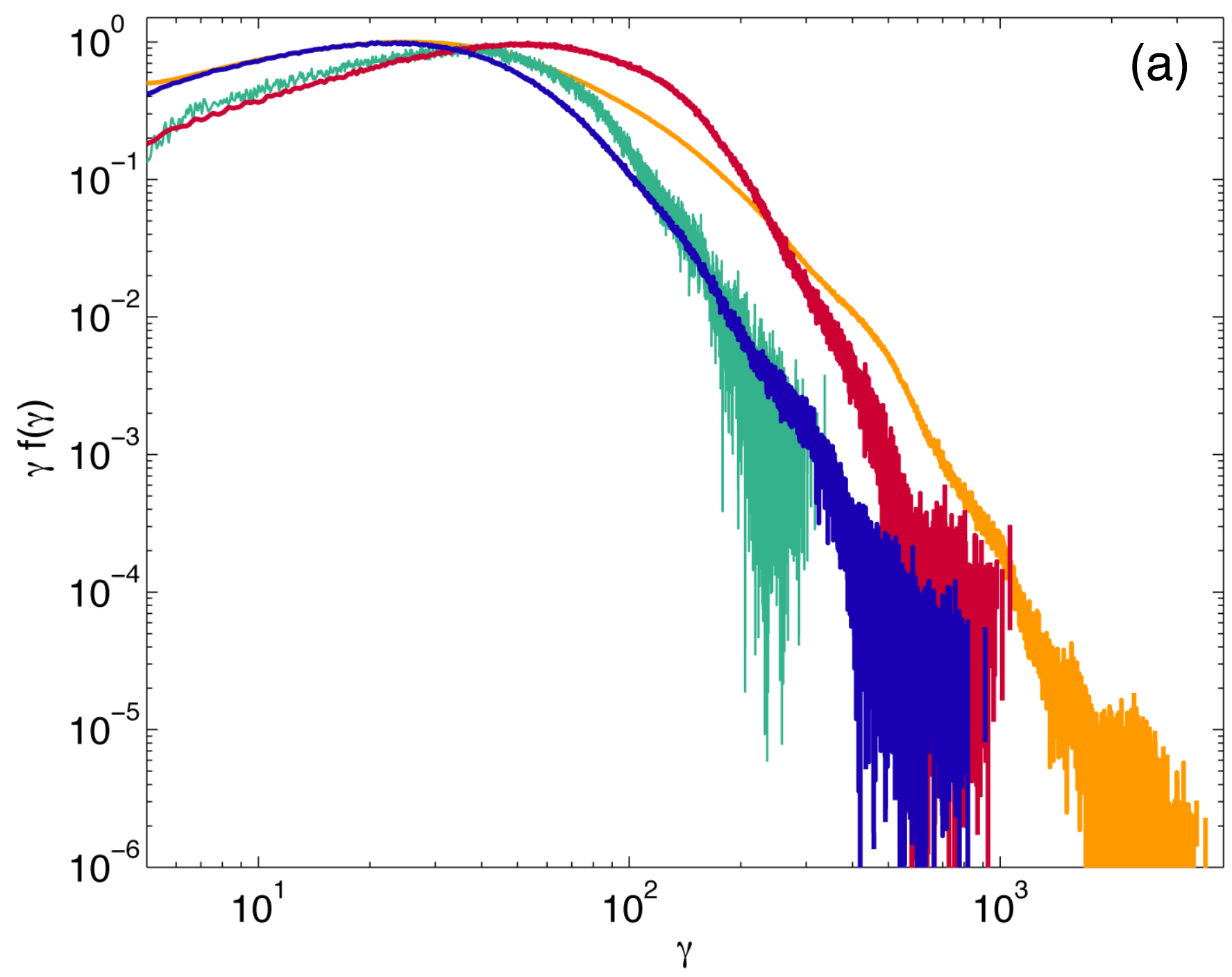}
\includegraphics[width=7cm]{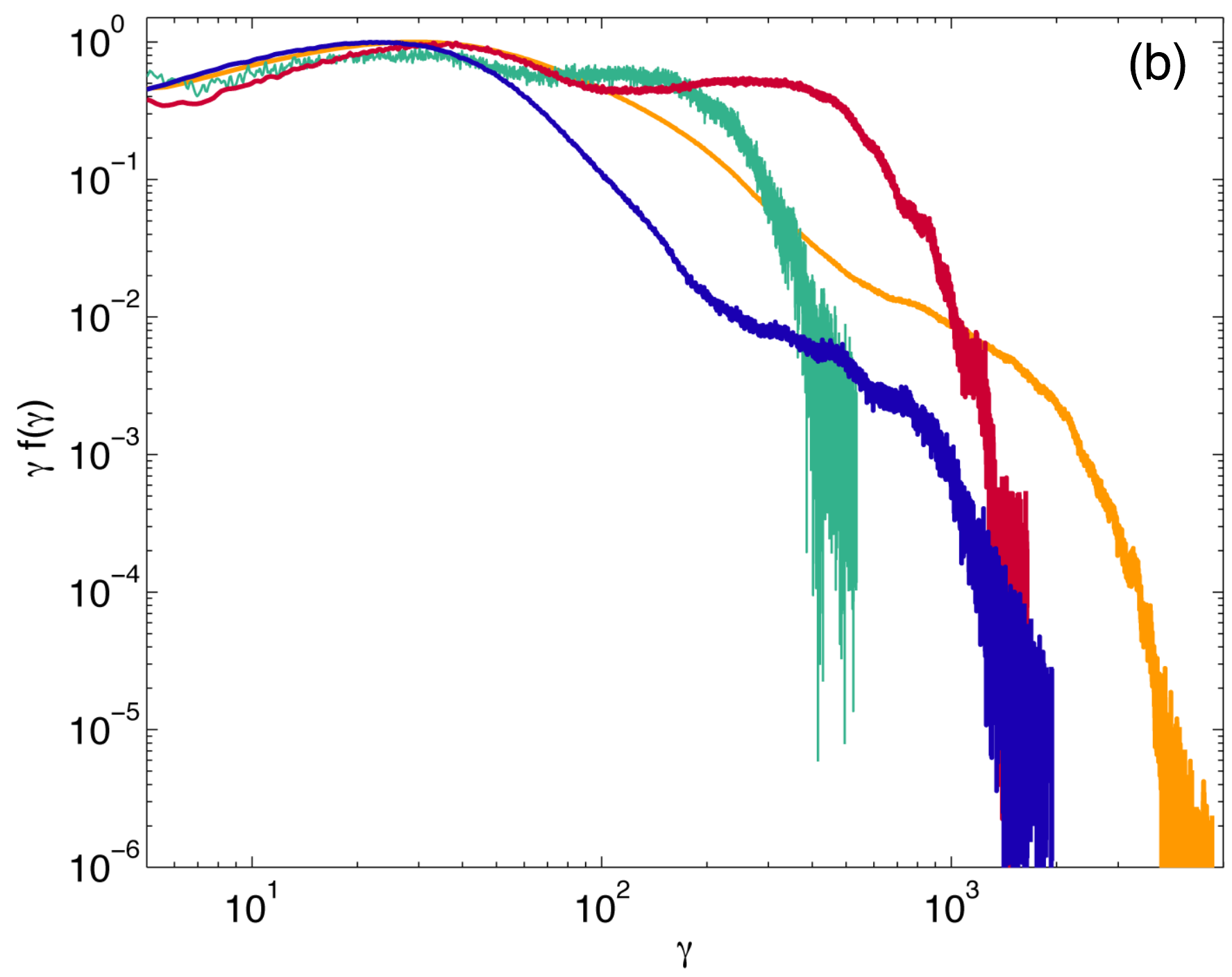}
\includegraphics[width=7cm]{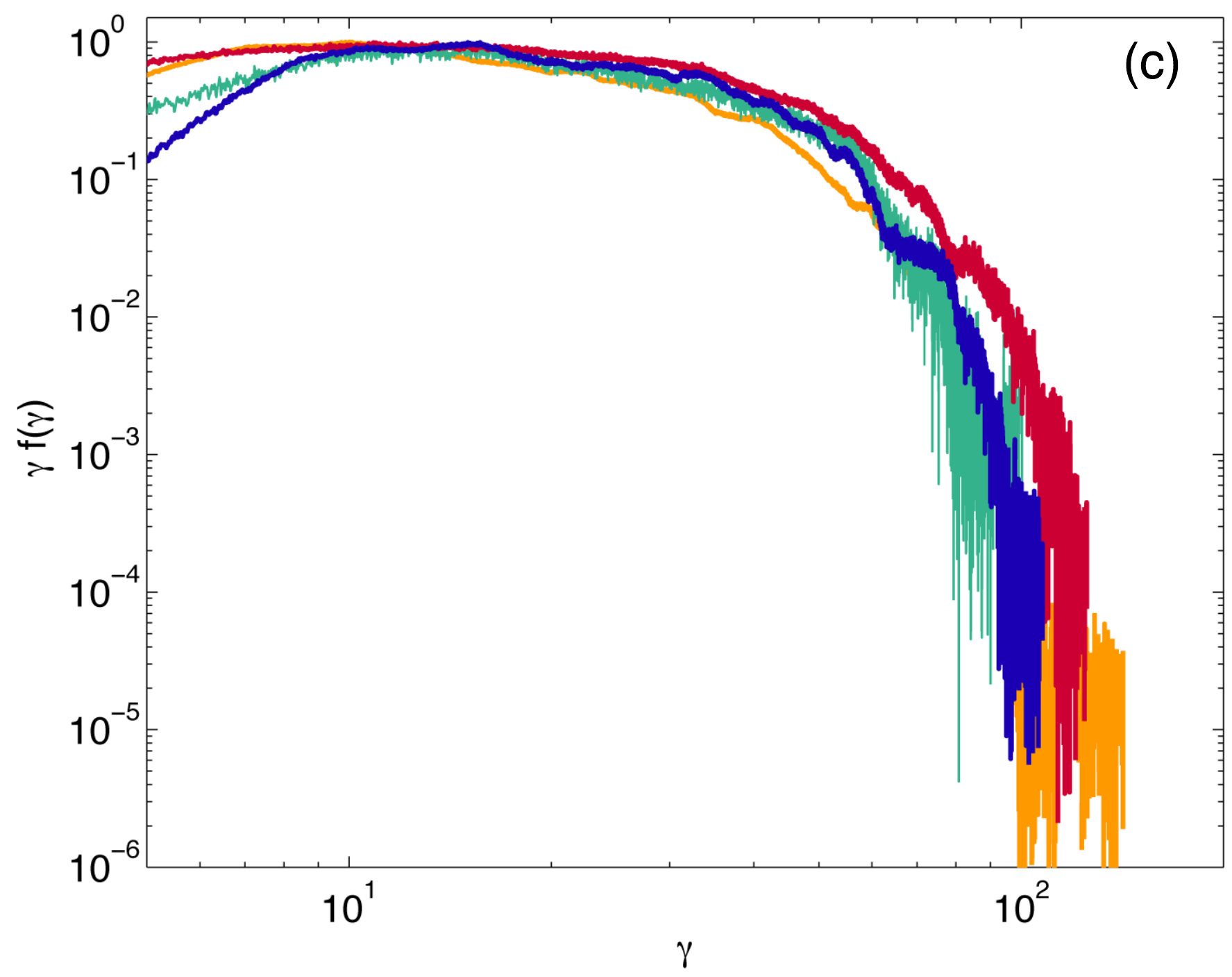}
\includegraphics[width=7cm]{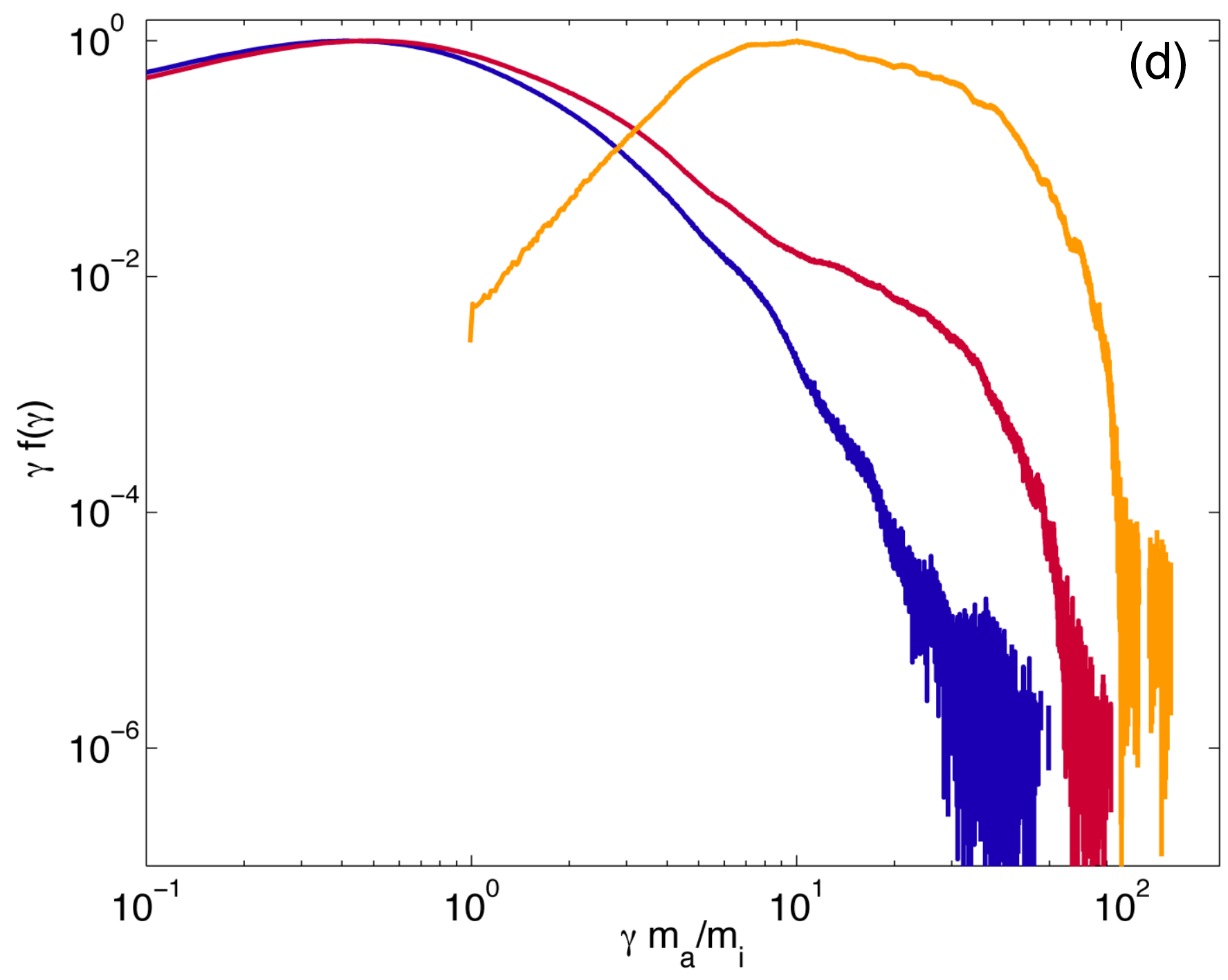}
\end{center}
\vspace{-18pt}
\caption{Electron (a), positron (b) and ion (c) downstream spectra. Color codings are given in Fig. \ref{tab:simpar}. (d) Comparison of electron (blue), positron (red) and ion (orange) spectra versus \( \gamma m_a / m_i\) with \(m_a / m_i\) the mass of the respective species normalized by the ion mass for \(\sigma_1 = 0.145\) and \(m_i/m_e= 60\). All spectra are plotted at \(t = 654 \, \omega_{ci}^{-1}\).}\label{fig:spectra}
\end{figure}

The comparison of the particle spectra is done at the same time in units of \(\omega_{ci}^{-1} = 6.3 \times 10^{-6} \, \textrm s^{-1} \times (m_p/m_e) / \sqrt{n_{12,e-} [\textrm{m}^{-3}]}\). Figure \ref{fig:spectra} shows the spectra at \(t = 654 \, \omega_{ci}^{-1}\). The electron spectra do not differ much from a thermal distribution, as well as the ion spectra, which have just thermalized, whereas the positron non-thermal tail is strong. A scaling of the maximum \(\gamma\) with the ion to electron mass ratio is apparent. For the density ratio \(n_{12,p+} / n_{12,e-} = 0.7\) the peak of the positron tail almost reaches the same level as the maximum of the thermal bulk and it is two orders of magnitude lower for \(n_{12,p+} / n_{12,e-} = 0.2\). The comparison of the distributions of the different species in Figure \ref{fig:spectra} (d) shows that the electron tail is much weaker than the positron tail. The ion spectrum has a completely different shape from the light species, which was also observed in the case of a pure electron-ion shock with high initial magnetization \cite[][]{SS11}.

\begin{figure}[ht!]
\begin{center}
\includegraphics{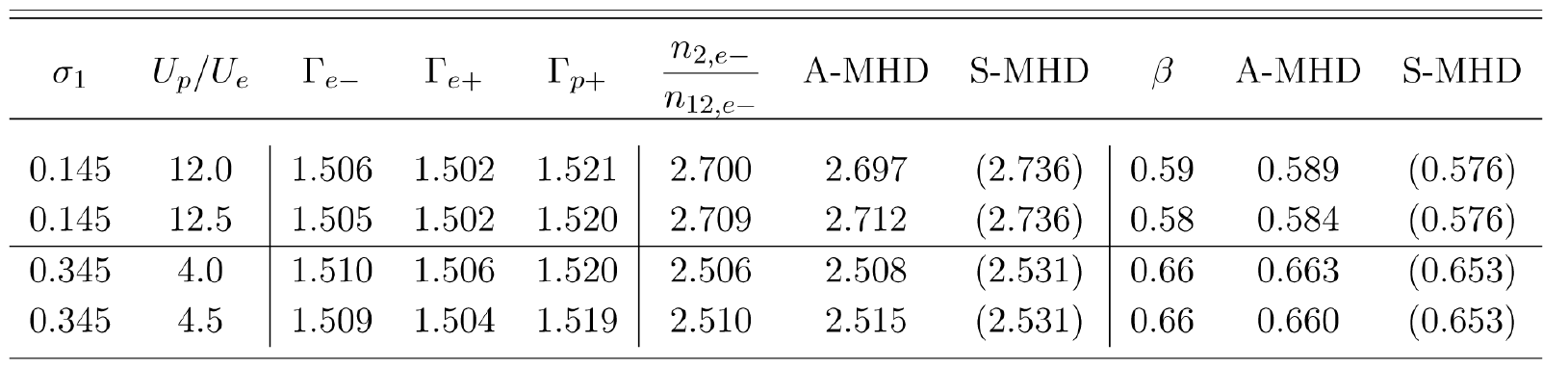}
\end{center}
\vspace{-18pt}
\caption{Downstream parameters measured from simulation data and comparison of density compression and shock speed with theory (A-MHD), obtained from Equations (\ref{exact}) and (\ref{densityjump}), and standard MHD theory (S-MHD) in brackets.}\label{tab1}
\end{figure}

The spectra are fitted with functions of the form given in Equation (\ref{fittingfunction}), with the parameters listed in Fig. \ref{tab:fit} in the Appendix, and used to calculate the jump conditions according to Equations (\ref{bineq}) and (\ref{densityjump}). In order to determine the adiabatic constant systematically, we also integrate the spectra numerically, with a deviation of the order of less than 0.1\% deviation from the analytical result. The jump conditions are determined with the standard MHD model (S-MHD), where the adiabatic constant is 3/2, and compared to the advanced model (A-MHD) given by Equations (\ref{bineq}) and (\ref{densityjump}). The comparison with the simulation data is given in Fig. \ref{tab1}. By plotting the transversely averaged density against \(x_1\) and \(t\) the velocity of the moving shock front is measured, which is almost perfectly constant after a few 100's of  \(\omega_{pe}^{-1}\). To determine the density jump, the density was averaged over the full downstream region. This value is also constant during the entire shock evolution.

From Fig. \ref{tab1} we see that the A-MHD model fits the simulation data better than the standard model, although the variations are only on the 1\% level. The dependence of the shock speed on the magnetization \(\sigma_1\) and ion kinetic energy \(U_p/U_e\) is in agreement with Figure \ref{fig:theoretical}. The variations for a constant total magnetization \(\sigma_1\) and different energy ratios are rather small, but also here the trend towards higher density compression ratios and lower shock speeds for increasing ion kinetic energy ratio \(U_p/U_e\) is clearly recognizable.

\subsubsection{Long time evolution of acceleration in mixed shock plasmas}

We also studied the temporal evolution of the adiabatic constant by integrating the spectra. The adiabatic constant of the positron spectrum initially decreases rapidly and is constant after a few \(\omega_{pi}^{-1}\). The adiabatic constant of the electron spectrum increases first and then slowly drops towards the lower limit \(3/2\) of an ideal two-dimensional gas. The ion constant seems to follow the same trend, and at the end of our simulations, it is still in the increasing stage.

\begin{figure}[ht!]
\begin{center}
\includegraphics{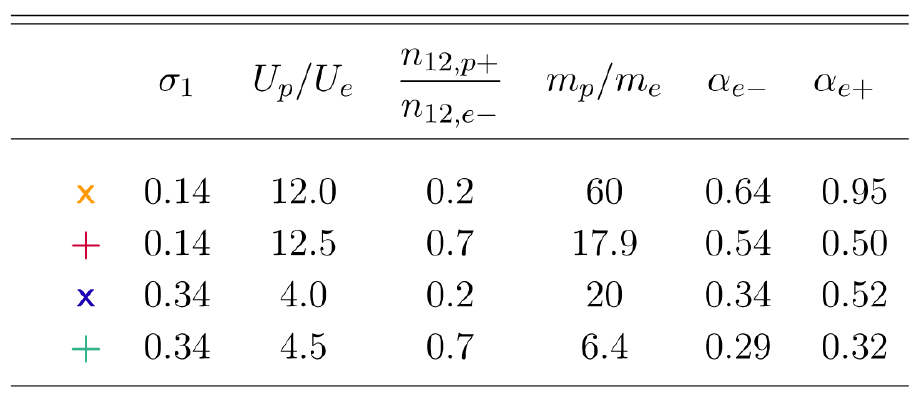}
\end{center}
\vspace{-18pt}
\caption{\(\alpha\) measured from the simulations according to \(\gamma_{max}  \propto (t-t_0)^\alpha\).}\label{tab3}
\end{figure}
\begin{figure}[ht!]
\begin{center}
\includegraphics[width=9cm]{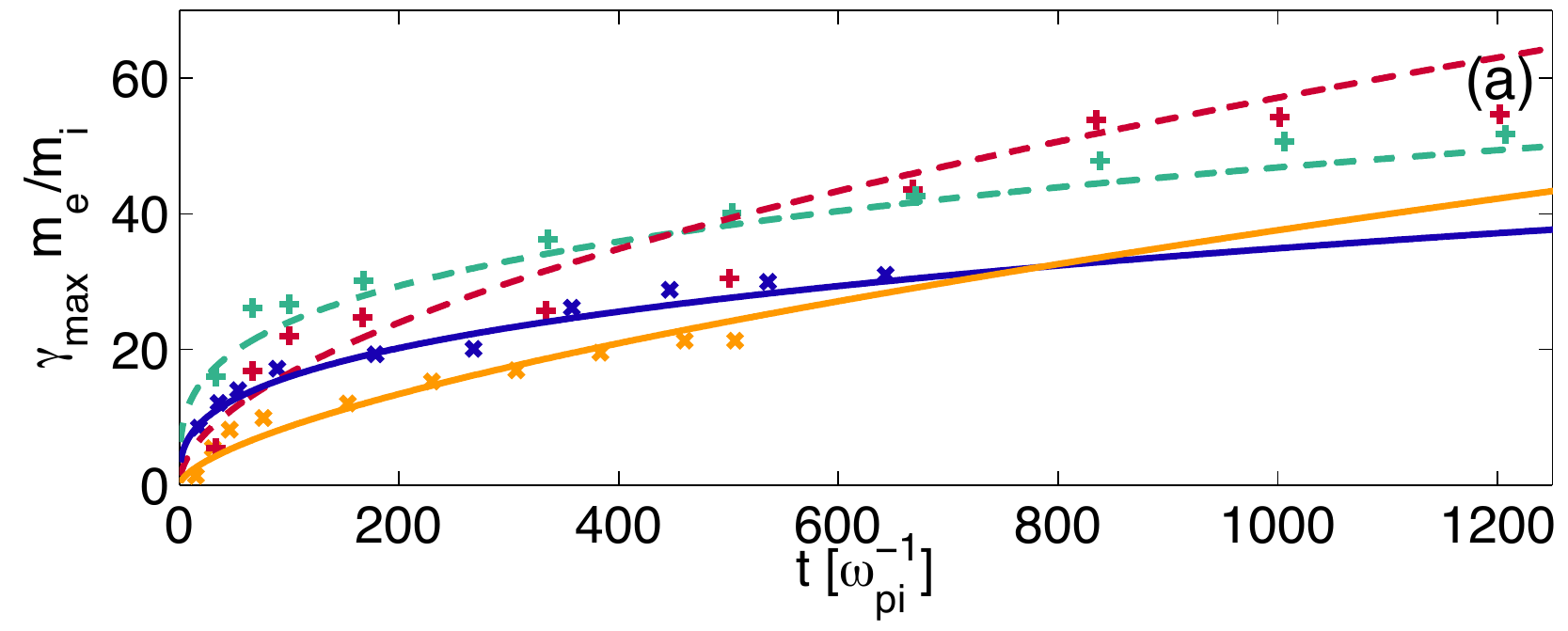}
\includegraphics[width=9cm]{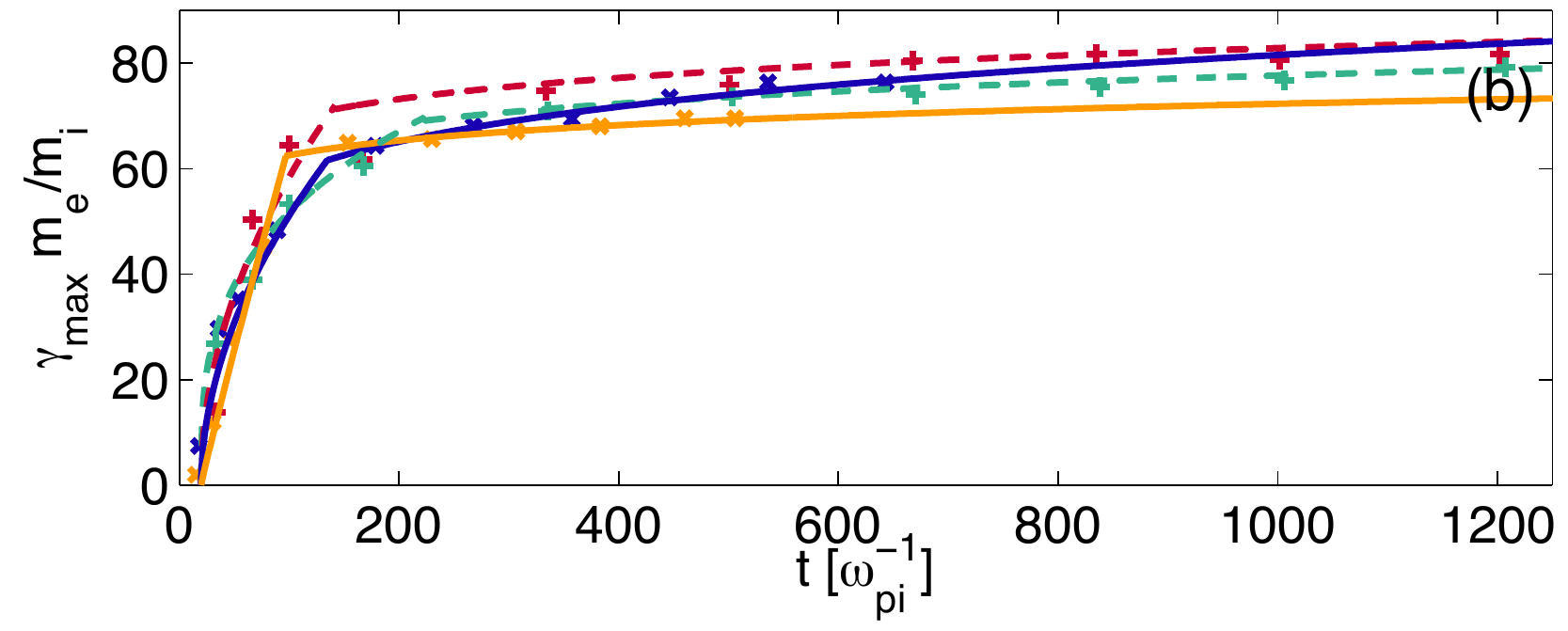}
\includegraphics[width=9cm]{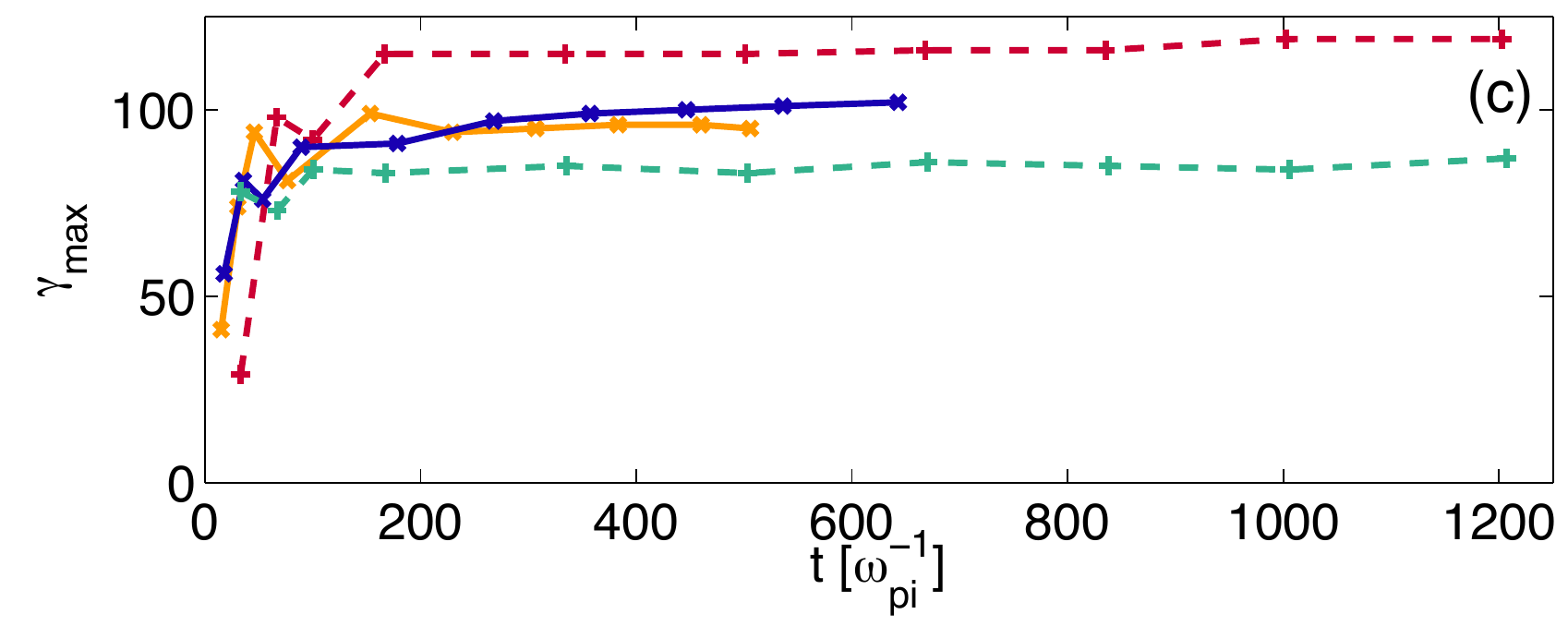}
\includegraphics[width=9cm]{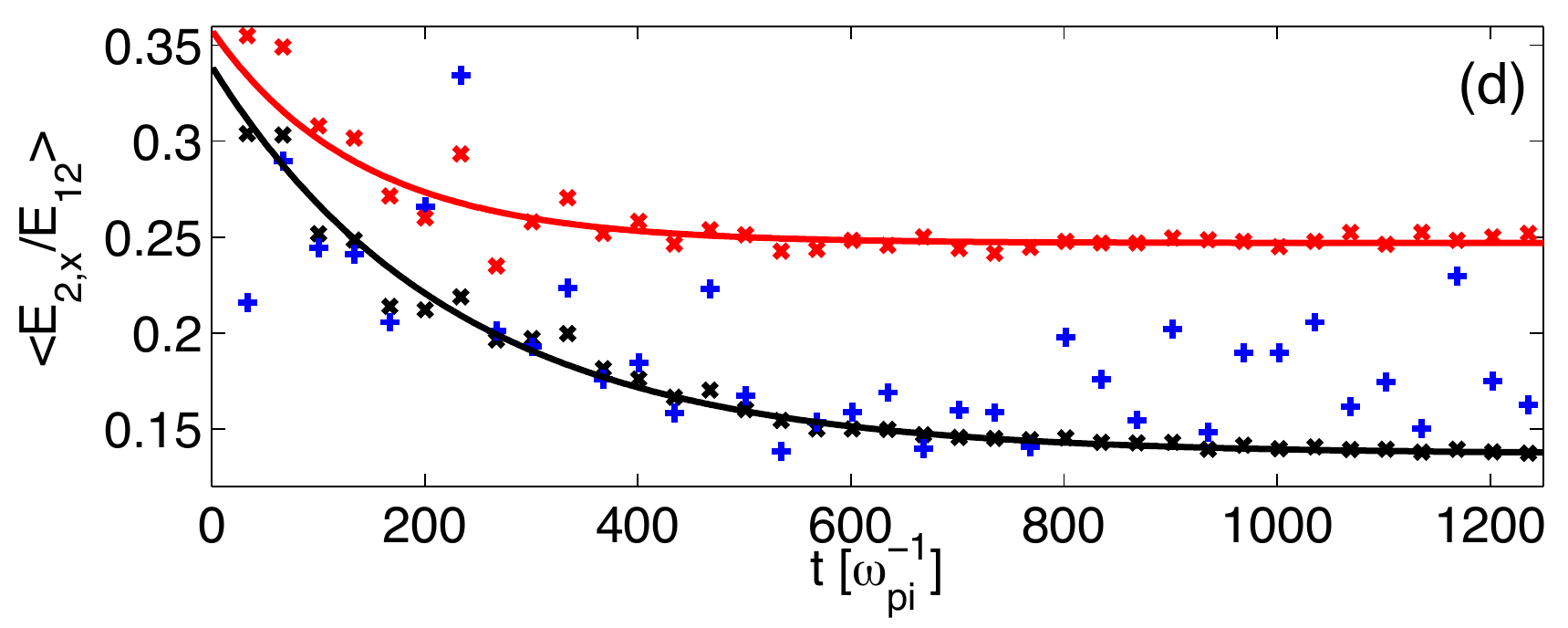}
\end{center}
\vspace{-18pt}
\caption{Maximum energy evolution for electrons (a), positrons (b) and ions (c) with fits to a power-law of the type \(\gamma_{max} \propto (t-t_0)^\alpha\) in (a) and (b) for electrons and positrons. The positron energies have been fitted with two power-laws. The index for low \(\gamma \) is given in Fig. \ref{tab3} together with the color codes. The index of the power-law for \(t> 200\, \omega_{pi}^{-1}\) is less than 0.1. (d) shows the evolution of the downstream electric field normalized to the initial upstream field averaged over the entire downstream region (black), in the far downstream (red) and in the shock front region (blue) over a range of \( 100 \, c/\omega_{pe}\) with exponentially decreasing fits.}\label{fig:gammamax}
\end{figure}

Figure \ref{fig:gammamax} (a)-(c) shows the temporal evolution of the maximum gamma of the non-thermal tail which finally determines the changes in the adiabatic constant.
We find, that the maximum energy scales like \(\gamma_{max}  \propto (t-t_0)^\alpha\), with the values for \(\alpha\) varying between \(\approx 1/3\) and 1, as shown in Fig. \ref{tab3}. The positron energy increases faster than the electron energy, which is in agreement with the preferred energy transfer due to the synchrotron maser instability. We observe a scaling with the ion mass ratio rather than with the kinetic energy ratio.
For a pure electron-ion shock in the unmagnetized case, a \(\alpha = 0.6\) has been observed (Fiuza et al. 2011, in preparation). The range of values for \(\alpha\) obtained here are consistent with acceleration due to multiple scattering in small wavelength turbulence (determined by the collisionless length/time scales) as predicted by \cite{KR10}.
For Bohm diffusion, the spatial diffusion coefficient scales like \(\kappa = \lambda v/3  \propto \gamma v^2\), estimating the acceleration time \cite[]{D83}
\begin{equation}
	 t_{acc} = \frac{3}{v_u-v_d} \int_{p_0}^{p} \frac{dp'}{p'} \left( \frac{\kappa_u(p')}{v_u} + \frac{\kappa_d(p')}{v_d} \right)
\end{equation}
as \(t_{acc} \propto \gamma\) \cite[]{GS11}, where \(v_u\), \(v_d\) are the upstream and downstream flow velocities and \(\kappa_u\), \(\kappa_d\) the upstream and downstream spatial diffusion coefficients.
According to \cite{KR10}, in the case of small-angle scattering the mean free path \(\lambda\) is rather proportional to \(\gamma^2\), as well as the spatial diffusion coefficient, and therefore the maximum energy is expected to evolve as \(t^{1/2}\) in the limit \(\gamma \gg 1\).

After \(t\geq 200 \, \omega_{pi}^{-1}\) the maximum positron energy stays almost constant (\(\alpha < 0.1\)). \cite{SS11} analyzed the acceleration mechanisms in pure electron-ion plasmas in an oblique magnetic field with magnetization \(\sigma_1 = 0.1\) and angle \(\theta= 75^\circ\) to the longitudinal component for a mass ratio \(m_p/m_e=16\) and found the synchrotron maser instability to be the dominant process in such a configuration. The transverse electromagnetic wave modes affect mostly the electrons, which leads to a decrease of their longitudinal momentum, whereas the heavy ions propagate almost with the initial momentum. The electrons are accelerated towards the shock by the induced longitudinal wakefield \cite[][]{L06} and it was observed that both species enter the shock region with almost the same energy, so that the electric field does not persist in the downstream. In contrast, we observe a non-zero electric field in the downstream region (Fig. \ref{fig:gammamax}d) which is clearly above noise level. During the process of ion thermalization, with the characteristic spiral structure in the \(p_2-p_1\) phase space \cite[][]{HA91}, the electric field is decreased due to the random motion of the ions, which causes the observed slowing down of the acceleration in Figure \ref{fig:gammamax}. The break in the power-law at  \(t\geq 200 \, \omega_{pi}^{-1}\) appears when the ions have finally thermalized. The electric field in the far downstream region has reached its asymptotic value \(0.25 \, |\mathbf E_{12}|\) at the same time. Moreover, at this stage, the positron energy becomes comparable to the ion energy, \(m_e \gamma_{2,e+} / (m_p \gamma_{2,p+}) \simeq 1 \), so that both positive species act in a similar way, whereas the electrons are still accelerated as their energy is small compared to the energy of the positive species.

Furthermore, \cite{SS11} observed a decrease of the electron tail to a thermal spectrum for \(t \omega_{pi} > 7000\) for the above mentioned setup. We have not seen any indication of such a decrease as even the positron spectrum remains almost constant. The detailed analysis of the long-term evolution and the influence of the field structure on the acceleration rate for these scenarios will be discussed in a future work.

\subsection{Decreasing the initial magnetic field}

We have performed a series of simulations with lower magnetizations than in the previous section in order to test the model for the jump conditions over a wide parameter range and compare them briefly to the case of a pure electron-positron plasma. The magnetic and electric fields are the same as in the previous setup. The two-dimensional simulation box consists of \(5000\, c / \omega_{pe} \times 50\, c / \omega_{pe} \) with cell size \(0.2 \, c/ \omega_{pe}\) and 6 particles per cell and species. We also did tests with up to 25 particles per cell, showing negligible differences. The two beams interact at \( x_1 = 300 \, c/ \omega_{pe}\), which allows us to fully resolve the shock dynamics, but also to reduce the box size. The time step is again chosen as \(\omega_{pe} \, \Delta t = 1/\sqrt{2}\) of the Courant condition in order to reduce simulation noise. In the run with \(m_p / m_e = 100\) the interaction region is shifted to \(x_1 = 1750 \, c/ \omega_{pe}\), and the simulation box is increased to \(10^4 \, c/ \omega_{pe}\) in propagation direction and the total simulation time is \(1.2 \times 10^4 \, \omega_{pe}^{-1}\). The different species configurations can be taken from Fig. \ref{tabrange} in the Appendix.

\subsubsection{Varying the ion fraction}

In the case of a pure pair plasma the electron and positron densities are nearly identical and show almost no spatial variation in the downstream. Only very weak filaments appear in the upstream region, which is an indication that the Weibel instability is almost completely suppressed. It is interesting to see that a magnetic precursor exists ahead of the shock also in the highly magnetized case. The phase spaces show a sharp transition between the downstream and the upstream region, also recognizable in the spatial density, which has been found to be a characteristic of superluminal shocks \cite[]{SS11}. The pair spectrum shows no evidence of a non-thermal tail and can be best fitted with \( f(\gamma) = C  \exp \left( -  \gamma/ \Delta \gamma \right)\) with \(\Delta \gamma = 17\).


\begin{figure}[ht!]
\begin{center}
\includegraphics[width=11.5cm]{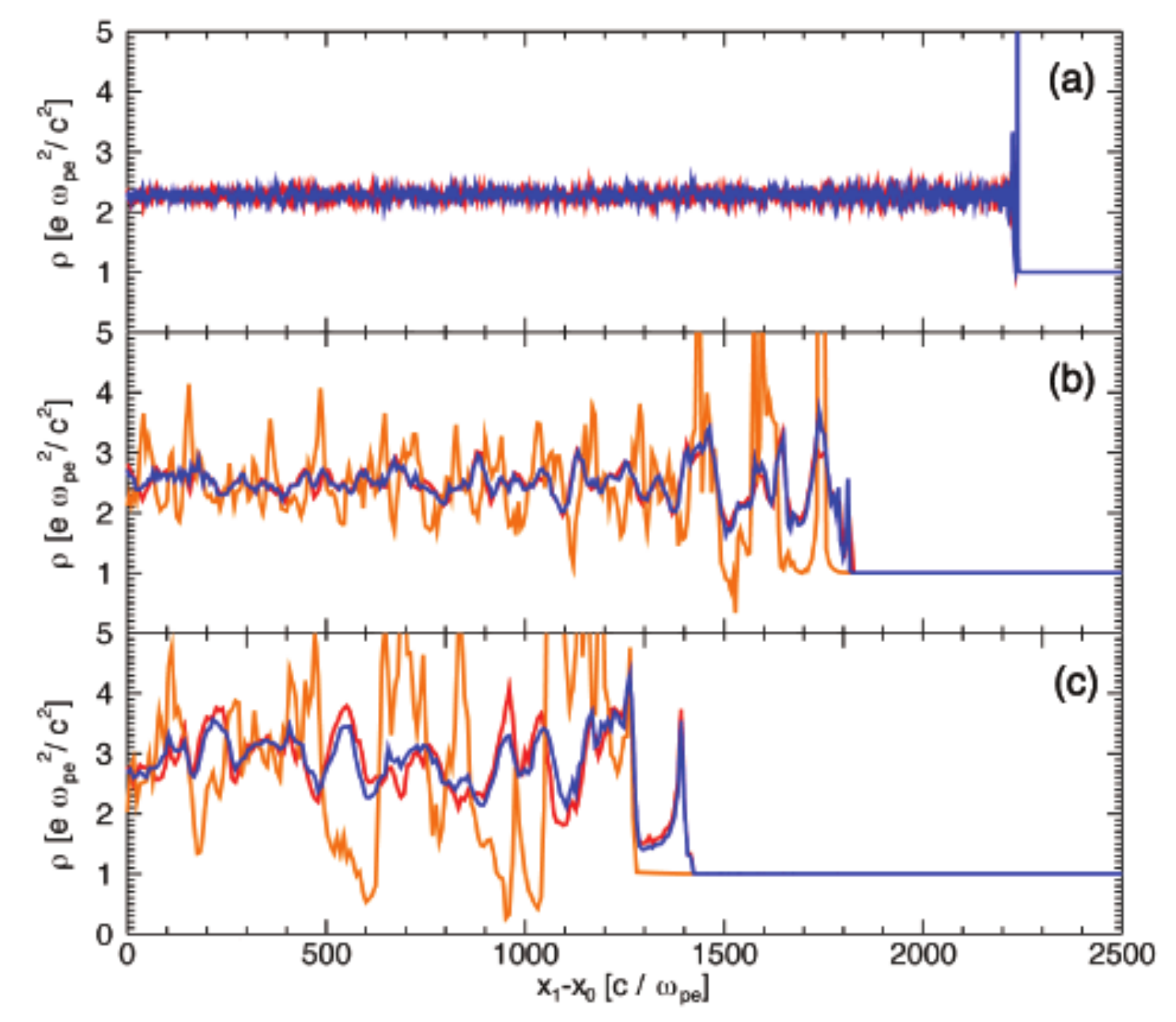}
\end{center}
\vspace{-12pt}
\caption{Comparison of the averaged and normalized densities \(n_{2,a}/n_{12,a}\) of electrons (blue), positrons (red) and ions (orange) with \(n_{12,p+} / n_{12,e-} =  0\) (a) and 0.2 else, \(m_p / m_e = 20\) (b) and \(m_p / m_e = 100\) (c). The simulation time is \(t \omega_{pe} = 3800\).}\label{fig:densitycomp}
\end{figure}

Moving from pure pair plasmas to mixed configurations, we observe that the shock speed is decreased and the density compression ratio increased if the ion to electron mass ratio is increased, demonstrated in Figure \ref{fig:densitycomp} for a density ratio \(n_{12,p+} / n_{12,e-} = 0.2\), which is in very good agreement with theory. The same behavior is found if the mass ratio is fixed and the density ratio is increased. The heavier ions cross the interaction region over a few electron skin depths, but soon they are reflected and the left-hand and right-hand populations are well-separated from each other. For demonstration purposes, the densities have been normalized to the upstream densities of each species, \(n_{2,a}/n_{12,a}\), showing that all three particle components behave similarly. The two-dimensional density profiles reveal a weak filamentary structure for electrons and positrons, like in the case of the pair plasma, which is not apparent in the ion density. Similarly to the unmagnetized case in \cite{SF11} the jump conditions only weakly depend on the real shape of the downstream spectra and the jump conditions in Fig. \ref{tabrange} are in good agreement with the theoretical model. The standard MHD model is a good approximation. Only at lower magnetizations we observe a stronger deviation from the simple model, in agreement with Figure \ref{fig:adiabat}. We observe that the final compression ratio is reached already after 3-4 \(\omega_{ci}^{-1}\), showing a steady state after 20-30 \(\omega_{ci}^{-1}\). \cite{H12} predict quasi-stationary solutions for low-\(\sigma\), but still not Weibel-governed, shocks only in case of electron-ion plasmas as the wave steepening will be stopped by energy dispersion into whistler waves, which are not present in pair plasmas.
Furthermore, if ions are present, the downstream structure along \(x_1\) shows strong wave generation on the scale of the ion Larmor radius. In the case of a pure electron-ion plasma, the structure is again much smoother and these oscillations become very weak, almost disappearing. The interaction between the three components is responsible for the oscillations present in the mixed component scenario, as previously observed in a 1D setup by \cite{HG92}. When the ions start to gyrate, the magnetic field is compressed, which leads to a compression of the pair density as well, because electrons and positrons are frozen into the field. The \(\nabla B\) drift generates a current that reacts back to the magnetic field, reinforcing the downstream compressional oscillations.
Indeed, we observe a perfect match between the out-of-plane magnetic field structure and the density of the pairs in the downstream. The ion density oscillates with the same frequency and a phase shift of \(180¡^\circ\), reaching maximum density when the pair density is minimum.

\begin{figure}[ht!]
\begin{center}
\includegraphics[width=8cm]{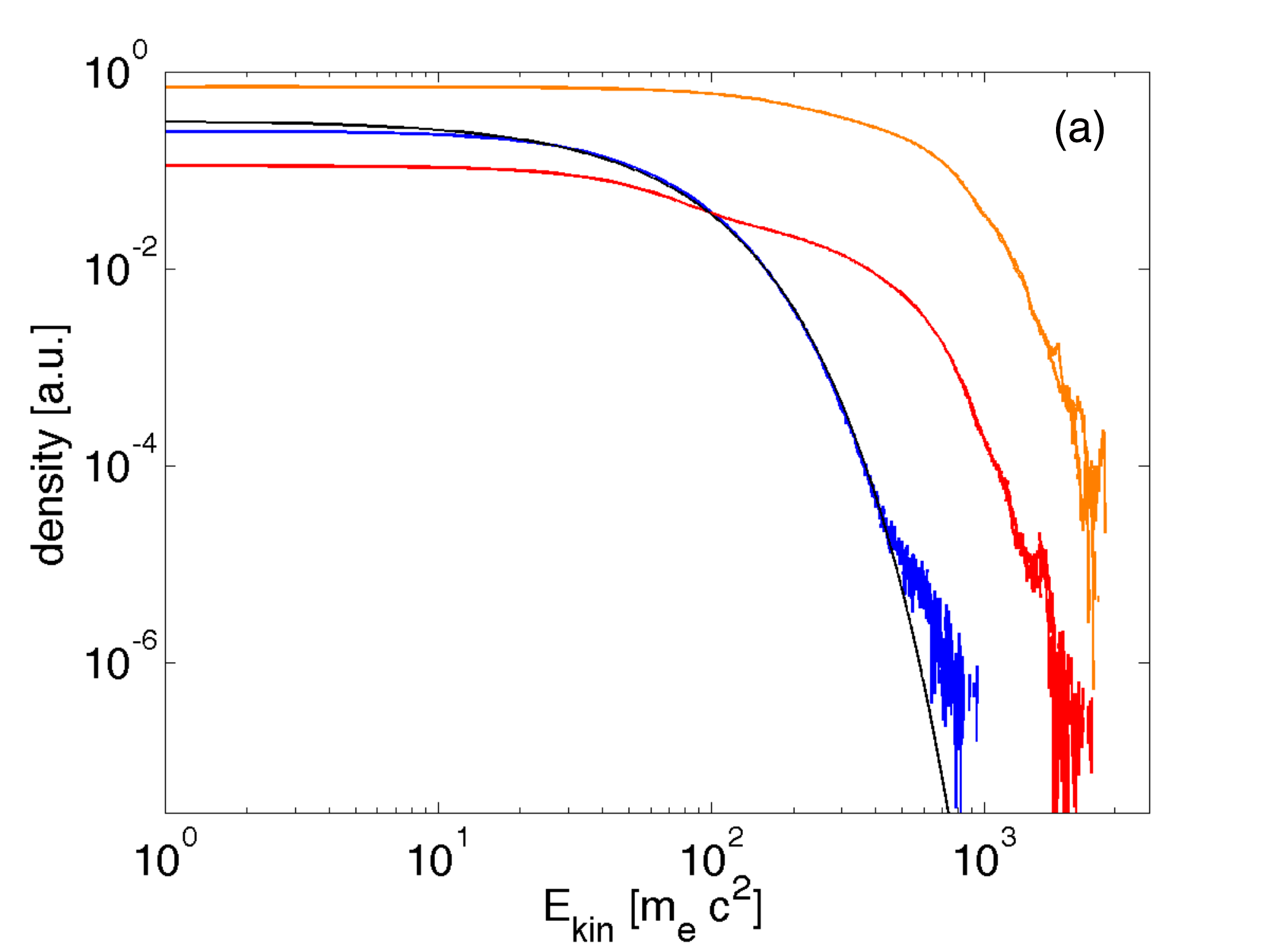}
\includegraphics[width=8cm]{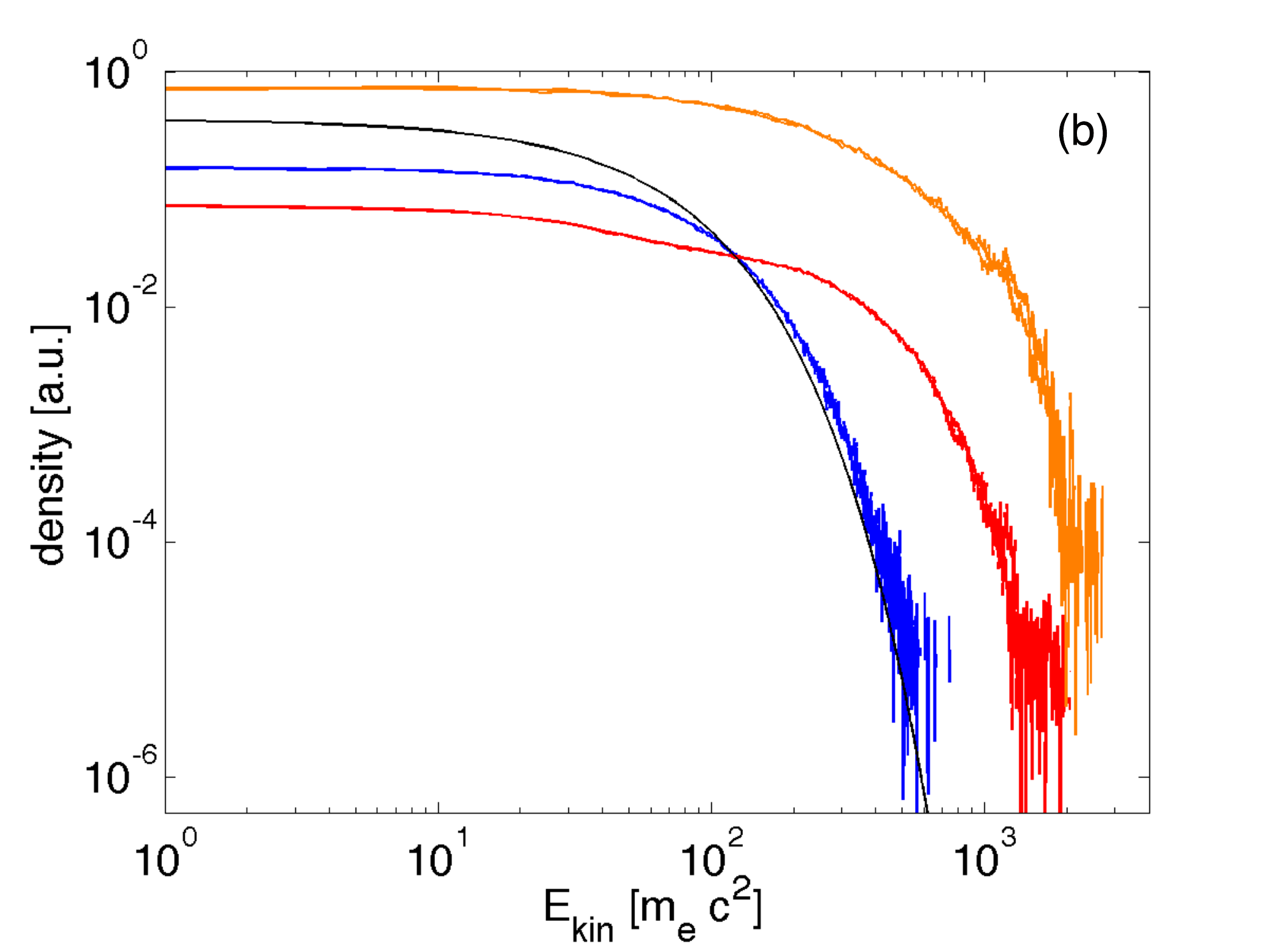}
\end{center}
\vspace{-12pt}
\caption{2D (a) and 1D (b) downstream distributions at \(t \omega_{pe} = 3800\) for positrons (red), electrons (blue), ions (orange) with \(m_{p} / m_{e} = 20\) and \(n_{12,p+}/n_{12,e-} = 0.6\) and Maxwellian fit to the 2D data in black.}\label{fig:m20fac601d2d}
\end{figure}

We observe that if the mass ratio is increased, the electron spectrum approaches the positron spectrum, which is due to the decreased total magnetization. But still in the run with a mass ratio of 100 and \(\sigma_1 < 10^{-2}\), the non-thermal electron tail stays weaker than the positron tail even for long simulation times.

The analysis of the densities and magnetic fields showed that spatial variations along \(x_2\) are low. Also the differences in the distribution functions, obtained from 1D and 2D simulations, are small (Figure \ref{fig:m20fac601d2d}), and arise essentially from the different statistics in 1D vs.\ 2D simulations, which justifies to study the effects of a realistic mass ratio in 1D simulations.


\subsubsection{Realistic mass ratio}

The ion mass ratio is further increased to a realistic proton to electron mass ratio \(m_p/m_e = 1836\), and the total magnetization is decreased to the limit \(\sigma \approx 10^{-3}\), where the Weibel instability starts to become important \cite[]{S05}. Since the shock formation is determined by the proton cyclotron time scale \(\omega_{cp+}^{-1}\), we can study this process in detail. On the one hand, the reduced geometry was chosen due to limited computational resources, on the other hand, Weibel modes are suppressed and can be excluded as the shock driving mechanism. Although we are slightly above the threshold, we are aware that 2D effects might become important and plan to investigate their role in future work.

Because of the large proton Larmor radius \(r_{Lp+} = 8200 \, c / \omega_{pe}\) for initial electron magnetization \(\sigma_e = 2\), we increase the one-dimensional box to \( 4 \times 10^4 \, c / \omega_{pe} \) with a cell size \(\Delta x_1= 0.25\, c / \omega_{pe} \), using 64 particles per cell and species. The interaction point of the counterpropagating beams is fixed at \(x_1 = 10^4 \,c / \omega_{pe}\) and the total simulation time is \(4 \times 10^4 \, \omega_{pe}^{-1} = 4.87 \, \omega_{cp+}^{-1}\) with the proton cyclotron time scale \(\omega_{cp+}^{-1} \approx r_{Lp+} / c\). Also in this case, the density ratios \(n_{12,p+}/n_{12,e-} = 0.2\), 0.6, 1 are investigated.

\begin{figure}[ht!]
\begin{center}
\includegraphics[width=8cm]{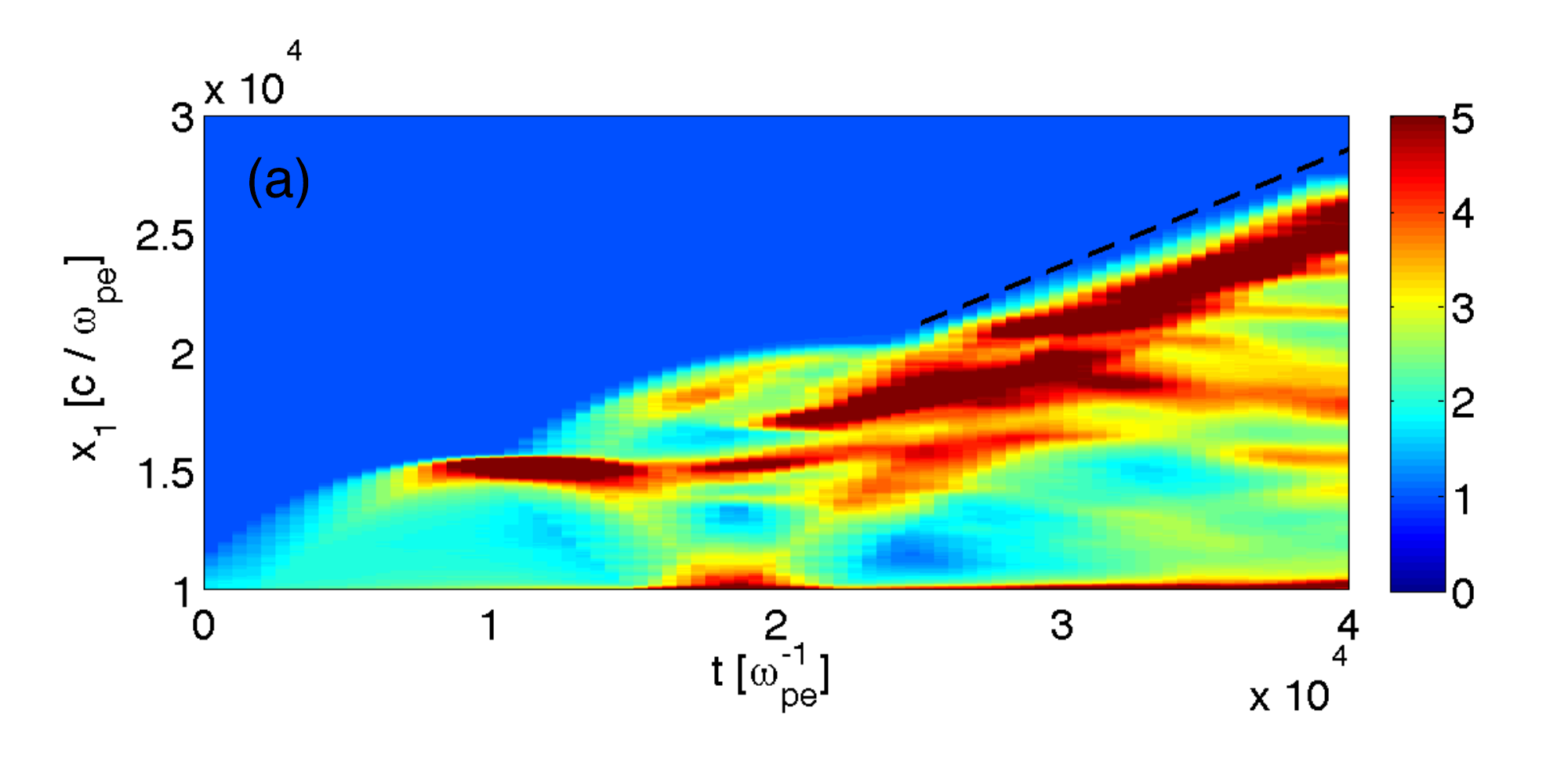}
\includegraphics[width=8cm]{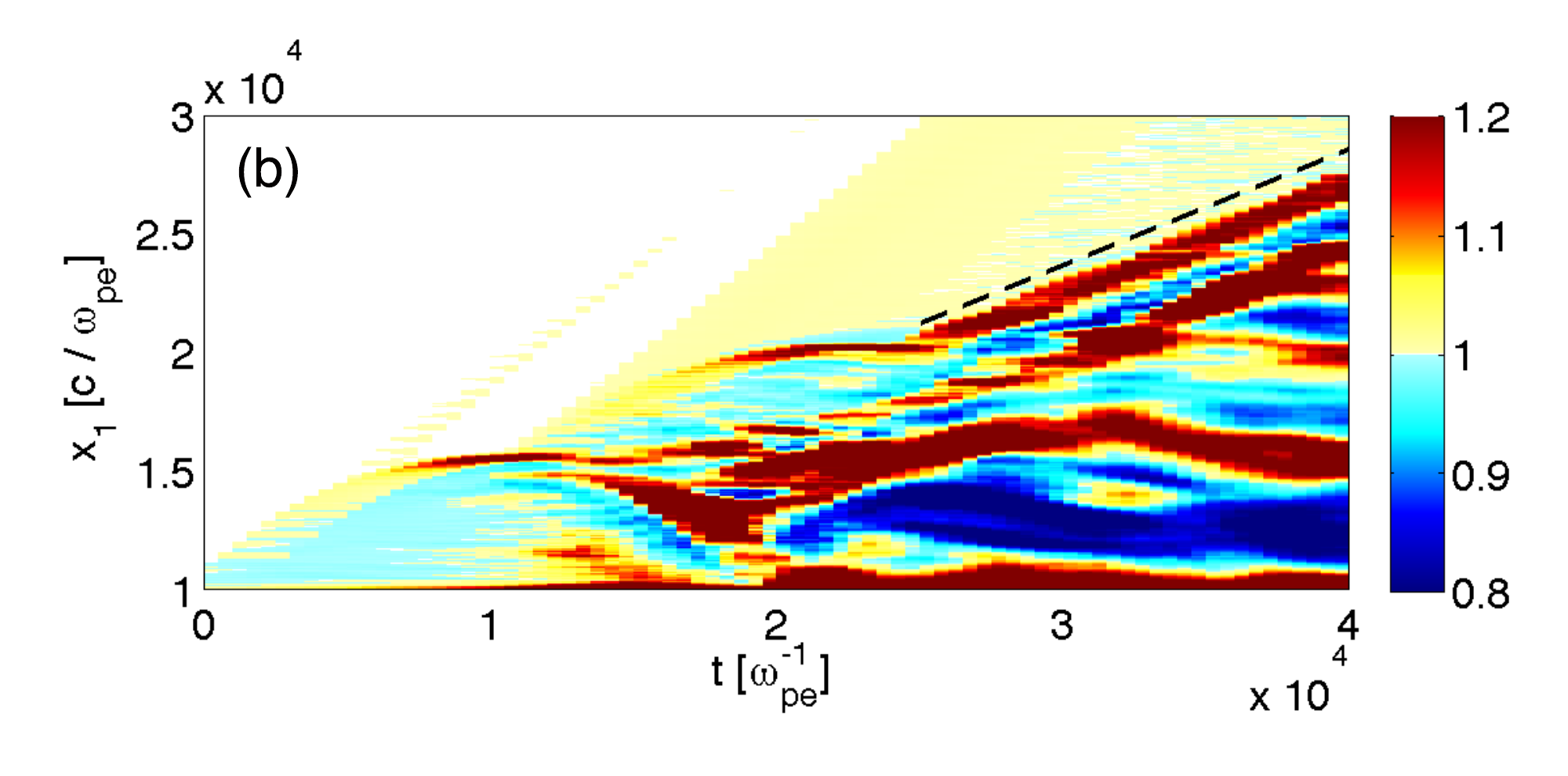}
\end{center}
\vspace{-12pt}
\caption{Electron density \(n_{2,e-}/n_{12,e-}\) (a) and electron to proton density ratio  \([n_{2,e-}/n_{12,e-}] / [n_{2,p+}/n_{12,p+}] \) (b) for realistic mass ratio \(m_{p} / m_{e} = 1836\) and \(n_{12,p+} / n_{12,e-} = 0.6\). The dashed lines indicate \(x_1 / t = 0.49\).}\label{fig:m1836fac60shock}
\end{figure}

The runs with a realistic mass ratio show a highly dynamic structure in the beginning, which appears to be almost independent of the initial ion fraction. Figure \ref{fig:m1836fac60shock} shows the electron density, for the case \(n_{12,p+} / n_{12,e-} = 0.6\), against \(x_1\) and \(t\) which allows the determination of the shock velocity \(\beta = x_1 / t\). 
Three stages have been identified from the analysis of the shock speed, which we discuss in detail for this density ratio.

In the first stage, which lasts until approximately \( 1.5 \, \omega_{cp+}^{-1}\), the ions are still cold and their phase space profiles differ much from that of the electrons and positrons. The counterpropagating beams are unaffected and propagate almost with the speed of light. At \(t \approx 4100 \, \omega_{pe}^{-1} = 0.5 \, \omega_{cp+}^{-1}\) the plasma is significantly compressed. At this stage we observe an extended ion gyro cycle, which reaches deeply into the downstream region, while the two populations of left and right electrons and positrons are almost separated and overlap only for a few electron gyro radii at the interaction region. The light particles are already thermalized and the beginning of a non-thermal profile is recognizable. While the electron distribution can be fitted well with a Maxwellian with thermal spread \(\Delta \gamma = 17\), the positrons already show a clear non-thermal tail.


In the second stage \( 1.5 \, \omega_{cp+}^{-1} < t < 3 \, \omega_{cp+}^{-1} \), the ions start to respond slowly to the generated magnetic field compressed at the shock front and their distribution deviates from the initial cold upstream distribution, but the population is still far from being thermalized. The ion average density shows a strong spread around the electron and positron profiles and waves are generated on the scale of the proton Larmor radius. Electrons and positrons are further accelerated by the cyclotron instability \cite[]{AA06}, with the strongest effect directly behind the shock front. In that region, the electron and positron spectra are almost equal, revealing a strong non-thermal component. The strong compression of the plasma particles directly behind the shock front is responsible for the rapid decrease of the shock velocity.
A large amount of particles is reflected with the speed of light to both sides of the shock fronts at \(t \approx 12300 \, \omega_{pe}^{-1} = 1.5 \, \omega_{cp+}^{-1}\) and the density structure in Figure  \ref{fig:m1836fac60shock} shows a second arc. 
The density profile becomes very dynamic, which has not been observed for low mass ratios and the phase spaces show large electron and positron momenta where there is a strong mixture of the ion population, e.\,g.\ for \(t \omega_{pe} = 23800\) at \(x_1 = 1.4 \times 10^4 \, c / \omega_{pe}\).

In the third stage, at approximately  \(t \approx 24600 \, \omega_{pe}^{-1} = 3 \, \omega_{cp+}^{-1}\), a quasi-steady structure is reached, where the shock speed matches the theoretical value \(\beta = 0.49\). But even for these simulation sizes it is difficult to clearly identify the formation of the shock. The density profile in the downstream region oscillates in space on the ion scale \(c/\omega_{ci}\), revealing regions of proton or electron accumulation, as observed in the previous section for reduced mass ratios. These regions appear to be quasi-steady in time (Figure \ref{fig:m1836fac60shock}).

The observed evolution of the shock speed is due to the involved dynamics of the different species. The electrostatic fields, which also appear in pure electron-ion shocks because of the different inertia, are only partially balanced, due to the presence of a light positive species. A precursor of electrons and positrons exists in front of the shock and the deceleration and acceleration of the light species resembles the crossing of the shock front in the Fermi acceleration process, which can enhance the acceleration of particles in mixed plasmas. 

\begin{figure}[ht!]
\begin{center}
\includegraphics[width=8cm]{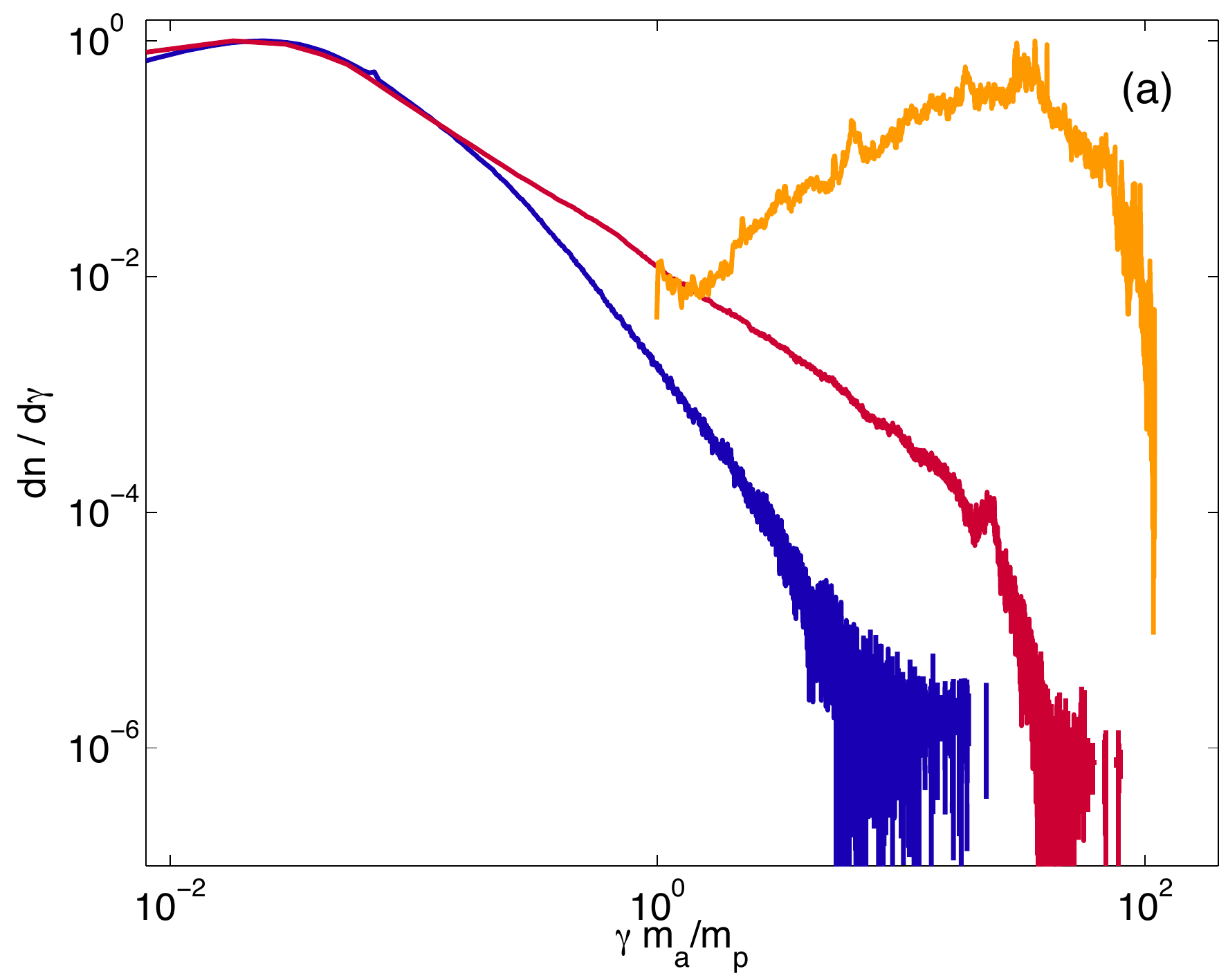}
\includegraphics[width=8cm]{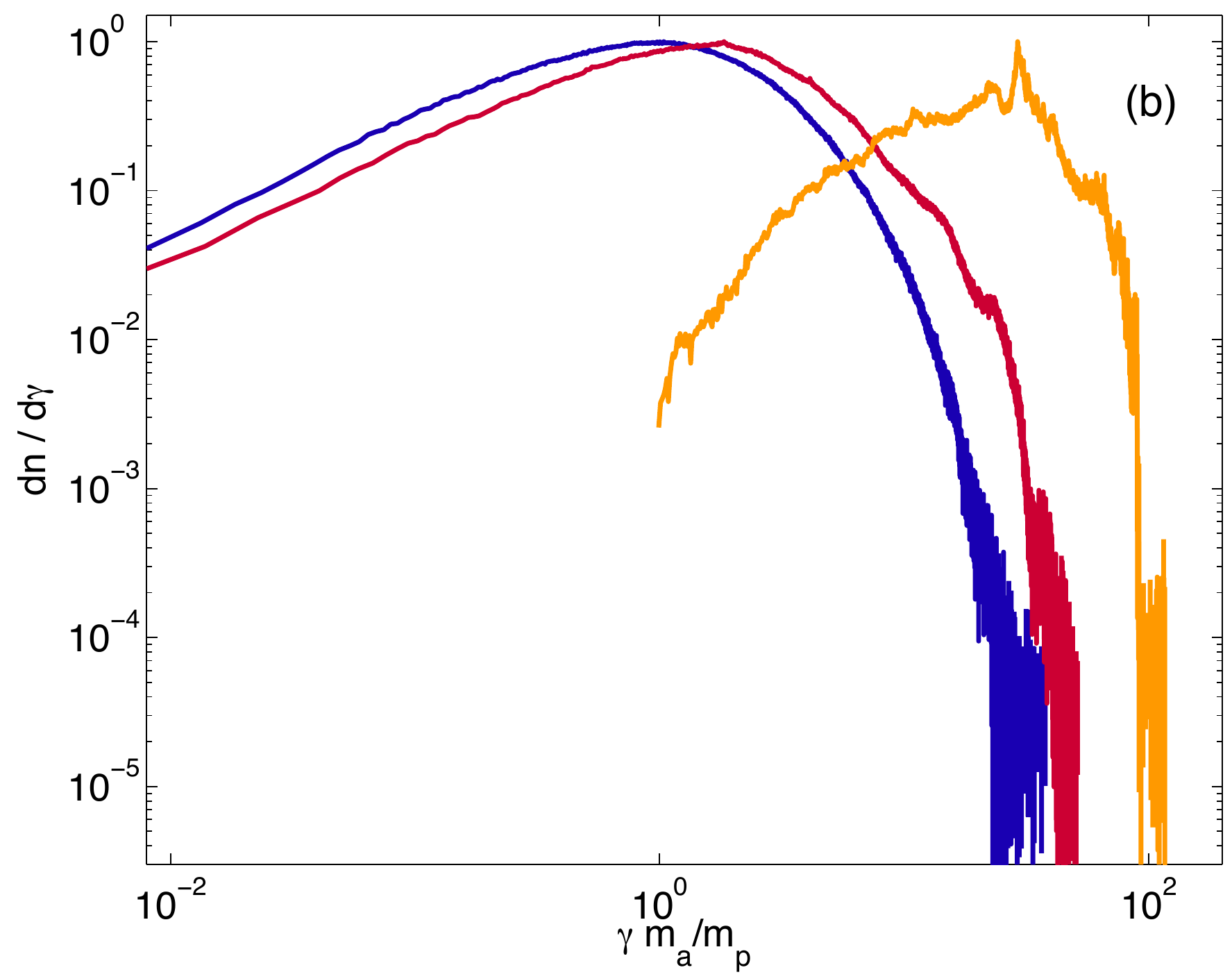}
\includegraphics[width=8cm]{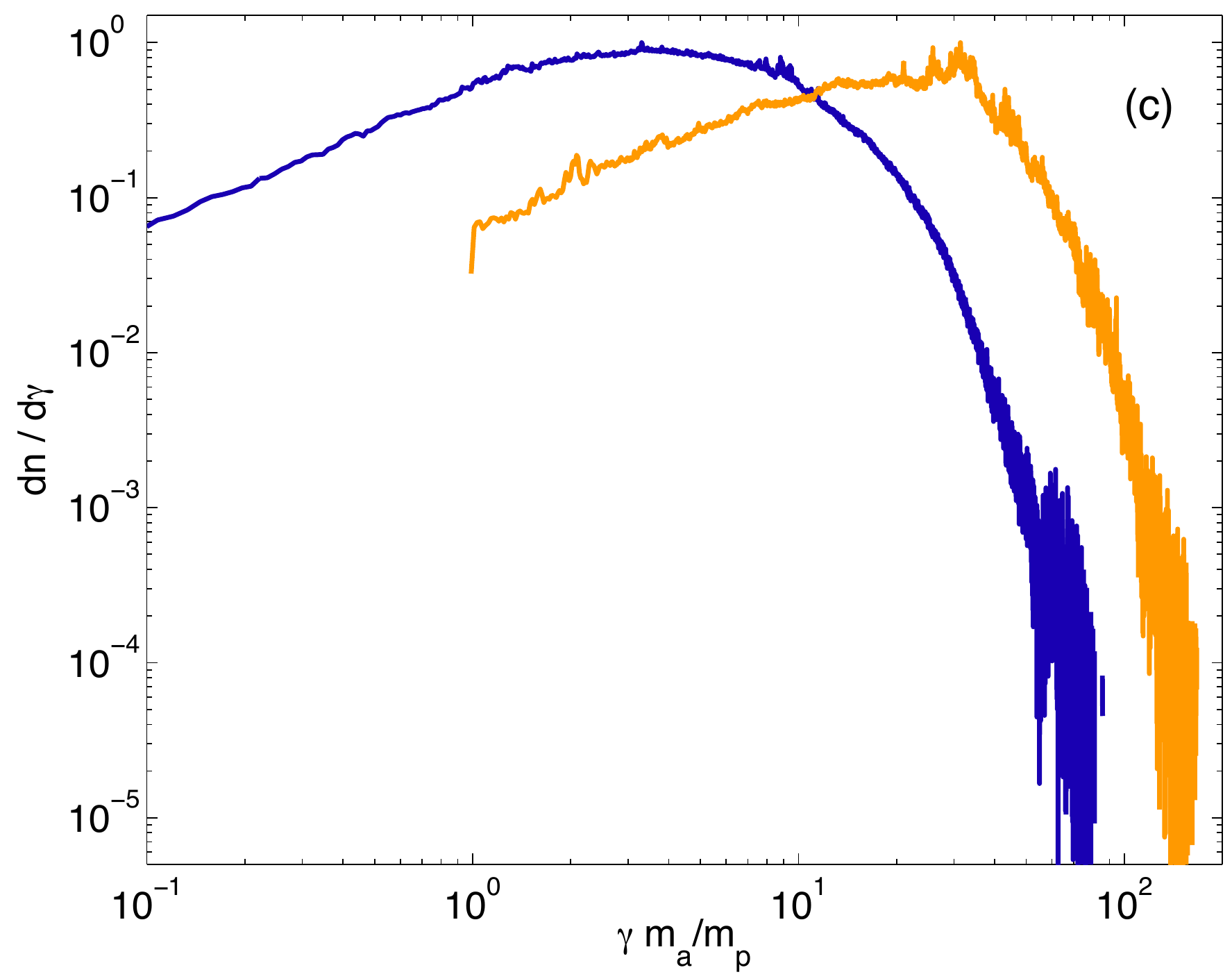}
\end{center}
\vspace{-18pt}
\caption{Electron (blue), positron (red) and proton (orange) distributions versus \(\gamma m_a / m_p\) with \( m_a / m_p\) the mass of the respective species normalized by the proton mass. The spectra have been averaged over the entire downstream region at \(t \omega_{pe} = 4 \times 10^4\). The initial density ratios are \(n_{12,p+} / n_{12,e-} = 0.2\) (a), 0.6 (b) and 1 (c).}\label{fig:m1836}
\end{figure}

In the other cases, \(n_{12,p+} / n_{12,e-} = 0.2\) and 1, we observe a similar qualitative behavior with the appearance of three temporal stages. In all three cases, the electron spectra show only weak non-thermal acceleration and deviate most from a thermal spectrum for a low proton to electron density ratio (see Fig.\ \ref{fig:m1836}a), but the spread \(\Delta \gamma\) of the peak energy of the thermal bulk increases with \( n_{12,p+}/n_{12,e-}\), as can be seen in Fig.\ \ref{fig:m1836}a-c, and the highest tail energies are achieved in the case of a pure electron-proton plasma (see Fig.\ \ref{fig:m1836}c). The maximum positron energy is independent of the proton fraction, but also here we observe an increase in the bulk spread. The proton spectra do not show evidence of non-thermal particle acceleration which is consistent with the results of \cite{SS11}.

\subsection{Magnetic field in the plane}

We performed simulations with the upstream magnetic field in the plane for the total magnetization \(\sigma_1 = 0.145\) and ion density ratio \(n_{12,p+}/n_{12,e-}= 0.7\). We compare the cases \(\theta= 90^\circ, \, 45^\circ, \, 0^\circ\) where \(\theta\) is the angle between the magnetic field and the longitudinal direction. The dominant acceleration process is determined by the magnetic field orientation, which was classified by \cite{SS11} into subluminal \(\theta < \theta_{crit} \simeq 34^\circ\) and superluminal shocks \(\theta > \theta_{crit}\). Accordingly, particles gain energy in subluminal shocks by non-resonant interactions with Bell's waves and are efficiently accelerated while bouncing back and forth across the shock. In superluminal shocks, if ions are present, the synchrotron maser waves transfer energy to the electrons. \cite{SS11} observed a short power-law tail stemming rather from heating than acceleration.

\begin{figure}[ht!]
\begin{center}
\includegraphics[width=8cm]{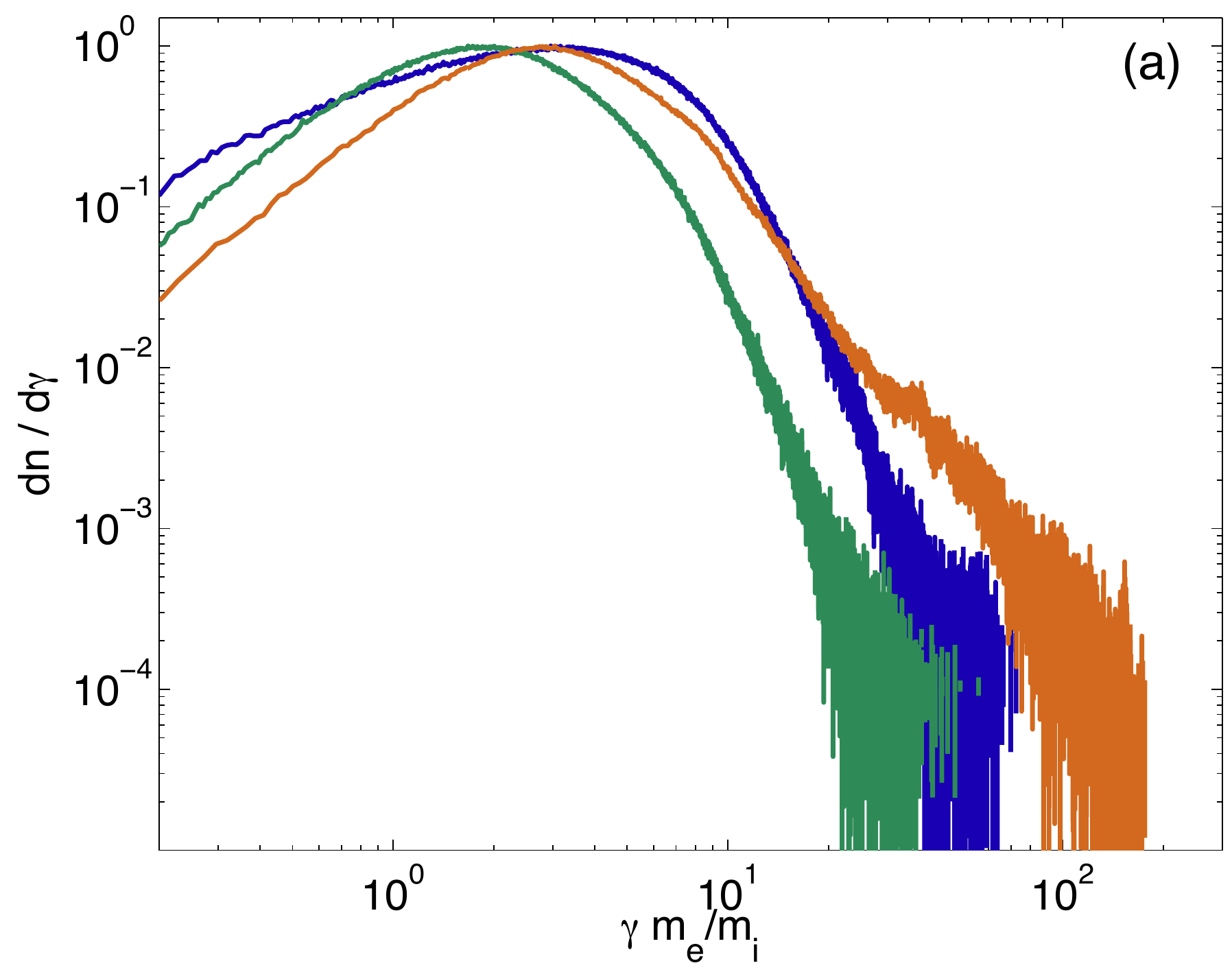}
\includegraphics[width=8cm]{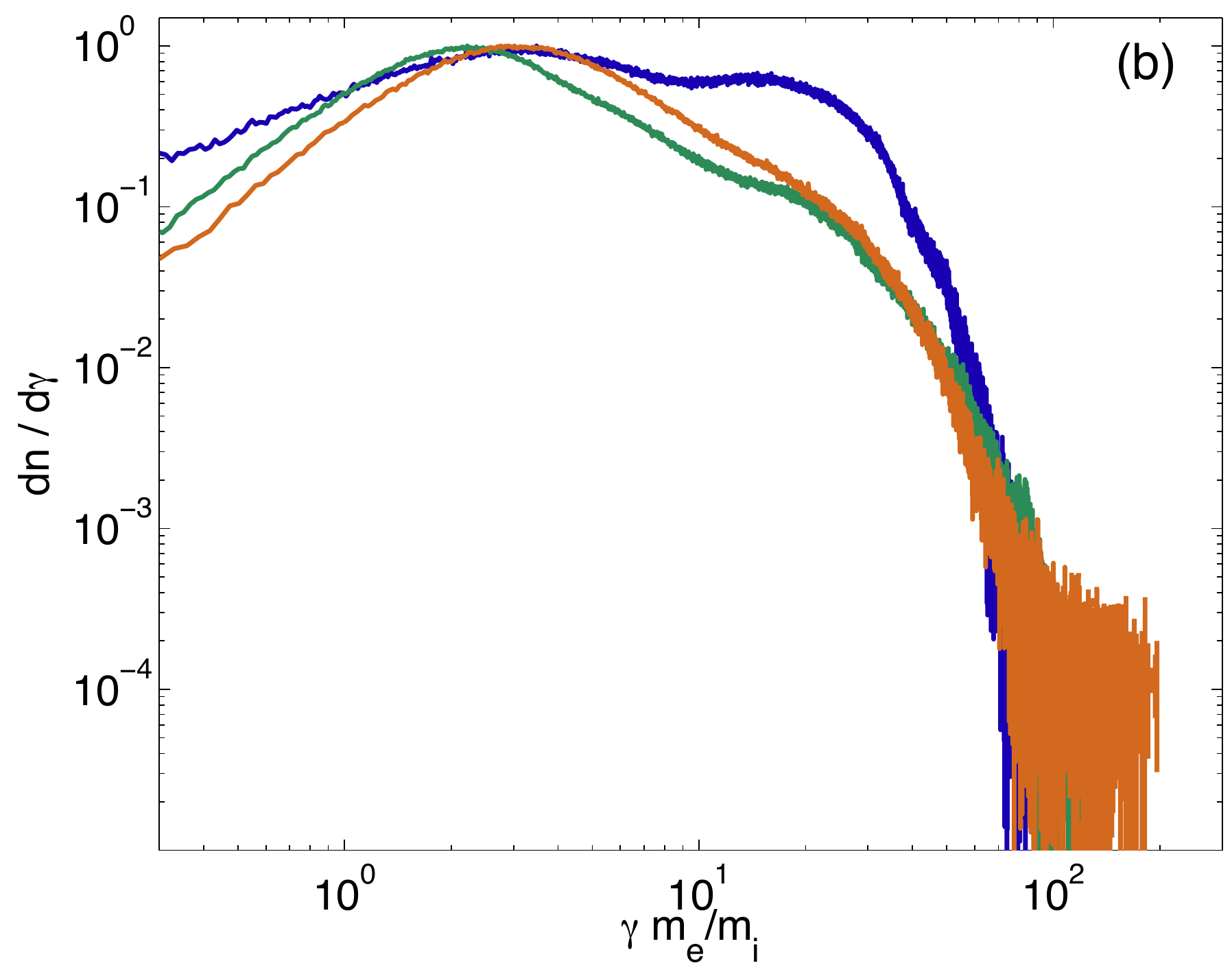}
\includegraphics[width=8cm]{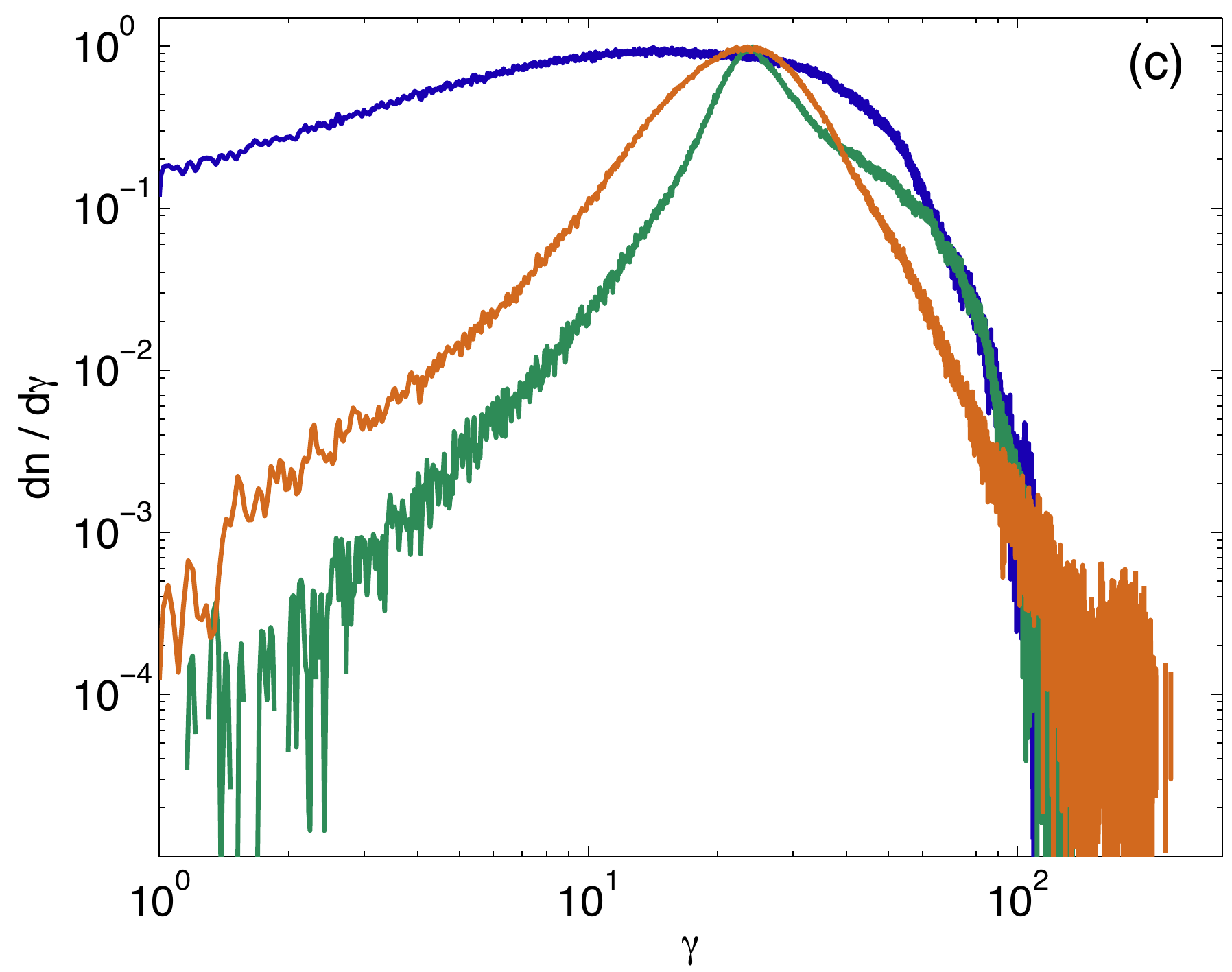}
\end{center}
\vspace{-18pt}
\caption{(a) Electron, (b) positron and (c) ion distributions for \(t \omega_{pe} = 6000\) and angle between upstream magnetic field and longitudinal direction \(\theta = 90^\circ\) (blue), \(\theta = 45^\circ\) (green), \(\theta = 0^\circ\) (brown).}\label{fig:perpparcomp}
\end{figure}

Our results are in agreement with these findings for the electrons. At the end of the simulation, at \(t \omega_{pe} = 6000\), the electron spectra show no sign of non-thermal acceleration if the magnetic field is superluminal, see Figure \ref{fig:perpparcomp}. In the subluminal case, the highest energies are achieved, which reach the level of that of the positrons. However, the positrons show a different behaviour. The maximum positron energy is the same in all three runs, with the strongest tail (=higher fraction of energy in the non-thermal particles) in the case of the perpendicular shock.

The ion spectra reach the same maximum kinetic energy as the positrons. The shape of the particle distributions varies much more than for the other components and resembles a thermal distribution only in the case of a perpendicular initial magnetic field. For the cases with a parallel component, the spectra become narrow, rather like a 3D-Maxwellian as it was observed for superluminal shocks also in \cite{SS11}. 

\section{Discussion}\label{sec:dis}

We have investigated the shock generation in plasmas consisting of electrons, positrons and protons for different perpendicular upstream magnetic fields. The standard one-fluid jump conditions have been extended for a multi-component plasma and the real shape of the downstream particle distributions has been taken into account. The calculations predict a decrease of the shock speed if either the mass ratio or the upstream ion fraction are increased. The higher the upstream magnetization, the less important become the effects of the real particle distribution.
If the deviations from a Maxwellian are low, or if the magnetization is high enough, for a highly relativistic upstream the shock speed is determined by a simple second order equation, which depends only on the effective upstream magnetization.

Simulations of shocks in mixed plasmas have been performed for a constant total magnetization with different initial ion kinetic energies and compared with our advanced theoretical model. The shock has been launched by injecting two counterpropagating beams from each side of the two-dimensional simulation box. Whereas the standard model predicts the same jump conditions independent of the ion kinetic energy, the advanced model fits better and predicts the increase in the compression ratio and decrease of the shock speed for increasing ion kinetic energy. Nevertheless, the differences are on the 1\% level only.

The evolution of the maximum energy was found to be a power-law with a power \(1/3 < \alpha < 1\), which depends on the initial parameters, but clearly indicates that the acceleration process is slower than the usually considered Bohm diffusion and is consistent with scattering off small-scale magnetic fluctuations. The non-thermal tail of the positrons was found to be constant after \(t \omega_{pi}^{-1} \approx 200\), which coincides with the thermalization of the ions. 

The theoretical model has been tested for a wide range of parameters, showing a good agreement between simulation and theory, but a weak dependence on the actual shape of the spectra. If the ion fraction is increased, the evolution of the density is similar in all components on time scales of the inverse ion cyclotron frequency. During this transition phase the light particles run ahead of the heavy ion species, undergoing an oscillation along the shock propagation direction. This separation of the species has been observed for a mixed plasma even in the case of no initial magnetization, whereas in a pure electron-ion plasma both species are always perfectly matched right from the beginning of the shock formation like in the case of a pure pair plasma.

We observed that two-dimensional effects become more important if the total magnetization is low due to the increased Larmor radius. Nevertheless, transverse spatial dependences were found to be low and the particle spectra are similar. 

\acknowledgments
This work was partially supported by the European Research Council (ERC-2010-AdG Grant 267841) and FCT (Portugal) grants SFRH/BPD/65008/2009, SFRH/BD/38952/2007, and PTDC/FIS/111720/2009. Simulations were performed at the IST cluster (Lisbon, Portugal).

\appendix

\section{Derivation of the jump conditions}\label{sec2}
The purpose of this section is to derive an expression for the shock speed in terms of known upstream quantities, with which the jump conditions for a scenario described in Figure \ref{fig:explain} can be determined. The following calculations base on the conservation equations for a single fluid in the paper by \cite{KC84}, which are expressed in the shock frame, where upstream and downstream components both propagate perpendicular to the shock front. The magnetic field is oriented perpendicular to the shock front, as well as to the propagation direction of the particles, as demonstrated in Figure \ref{fig:explain}. The conservation equations in the shock frame for multiple species are thus given by
\begin{eqnarray} 
	n_{1,a} u_{1s} & = & n_{2,a} u_{2s}  \label{Jump_shock1}\\
	 \beta_{1s} B_{1s} & =  & \beta_{2s} B_{2s}  \label{Jump_shock2}\\
	\beta_{1s} \gamma_{1s}^2 \sum_a n_{1,a} \mu_{1,a} + \frac{\beta_{1s} B_{1s}^2}{4 \pi } &=& \beta_{2s} \gamma_{2s}^2 \sum_a n_{2,a} \mu_{2,a} + \frac{\beta_{2s} B_{2s}^2}{4 \pi }  \label{Jump_shock3} \\
	u_{1s}^2 \sum_a n_{1,a} \mu_{1,a} + \sum_a p_{1,a} + \frac{B_{1s}^2}{8 \pi}   & = & u_{2s}^2 \sum_a n_{2,a} \mu_{2,a} + \sum_a p_{2,a} + \frac{B_{2s}^2}{8 \pi}  \label{Jump_shock4}
\end{eqnarray}
with proper velocity \(u_{is} = \beta_{is} \gamma_{is}\), perpendicular magnetic field component \(B_{is}\), specific enthalpy \( \mu_{i,a} = m_a c^2 + (e_{i,a} + p_{i,a} ) / n_{i,a}\), rest frame energy density \(e_{i,a}\) and pressure density \(p_{i,a}\). In the following, we assume a cold upstream with \( p_{1,a} = e_{1,a} = 0\). Similar to the case of a single fluid, a downstream adiabatic constant \(\Gamma_a\) can be defined for each species by the pressure-energy relation \(p_{i,a} = (\Gamma_{a} -1 ) (e_{i,a} - n_{i,a} m_a c^2)\). Thus, the sum
\begin{equation}
	\sum_a n_{i,a} \mu_{i,a} =\sum_a n_{i,a} \left\{  m_a c^2  + \frac{\Gamma_{i,a} }{\Gamma_{i,a} -1}\frac{ p_{i,a}}{n_{i,a}} \right\}
\end{equation}
can be expressed by
\(\sum_a n_{1,a} \mu_{1,a} =: n_{1,e-} w_1 m_e c^2\) and
\( \displaystyle \sum_a n_{2,a} \mu_{2,a} =: n_{2,e-}  \left( w_2 m_e c^2 + w_3  \frac{\Gamma_{e-} }{\Gamma_{e-} -1}\frac{ p_{2,e-}}{n_{2,e-}}  \right)\) with
\(\displaystyle w_1 = 1 + \frac{n_{1,e+}}{n_{1,e-}} + \frac{n_{1,p+}}{n_{1,e-}} \frac{m_p}{m_e}\),
 \(\displaystyle w_2 = 1 + \frac{n_{2,e+}}{n_{2,e-}} + \frac{n_{2,p+}}{n_{2,e-}} \frac{m_p}{m_e}\),
 \(\displaystyle w_3 = 1 + \frac{\Gamma_{e+}(\Gamma_{e-}-1)}{\Gamma_{e-}(\Gamma_{e+}-1)}  \frac{p_{2,e+}}{p_{2,e-}} + \frac{\Gamma_{p+}(\Gamma_{e-}-1)}{\Gamma_{e-}(\Gamma_{p+}-1)}  \frac{p_{2,p+}}{p_{2,e-}}  \).
Combining Equations (\ref{Jump_shock1})-(\ref{Jump_shock4}) yields the determination equation of the shock speed in the shock frame, given by
\begin{equation}
 \frac{\gamma_{1s}}{\gamma_{2s}} \left[1+ (1-Y) \sigma_1\right] \left\{ 1+ \frac{w_3}{w_4} \frac{\Gamma_{e-}}{\Gamma_{e-}-1} u_{2s}^2 \right\} - \frac{w_2}{w_1} - \frac{w_3}{w_4} \frac{\Gamma_{e-}}{\Gamma_{e-}-1} u_{1s} u_{2s} \left[ 1+ \frac{\sigma_1}{2\beta_{1s}^2} (1-Y^2) \right] = 0
\end{equation}
with \(Y := \beta_{1s} / \beta_{2s}\) and \(\displaystyle w_4: = 1+  \frac{p_{2,e+}}{p_{2,e-}}  +  \frac{p_{2,p+}}{p_{2,e-}} \). The total magnetization is defined as
\begin{equation}\label{def:sigmatot}
	 \frac{ 4 \pi \sum_a n_{i,a} \mu_{i,a} \gamma_{ij}^2}{B_{ij}^2} :=\frac{ 1}{\sigma_{i} }=\frac{ 1}{\sigma_{i,e-} } + \frac{ 1}{\sigma_{i,e+} } + \frac{ 1}{\sigma_{i,p+} }
\end{equation}
according to \cite{HG92}, providing the total upstream magnetization \(\sigma_1 = B_{12}^2 / ( 4 \pi n_{12,e-} w_1 m_e c^2\gamma_{12})\). Performing a Lorentz transformation into the downstream frame, using
\(n_{i,a} = n_{ij,a}/\gamma_{ij} \),
\(B_i = B_{ij} / \gamma_{ij}\),
\(\beta := \beta_{s2} = - \beta_{2s} >0\),
\(\gamma:= \gamma_{2s} = \gamma_{s2} \),
\(\beta_{1s} = - (|\beta_{12}| + \beta)/(1+ |\beta_{12}| \beta)\) and
\(\gamma_{1s} = \gamma \gamma_{12} (1+ \beta  |\beta_{12}|)\)
one obtains the determination equation for the shock speed with parameters defined in the downstream (\(\, \hat= \,\)simulation) frame
\begin{eqnarray}\label{app:exact}
 && \gamma_{12}  \left[ 1+ \beta | \beta_{12} | - \frac{ |\beta_{12} |}{\beta \gamma^2} \sigma_1 \right] \left[ 1+  \frac{w_3}{w_4} \frac{\Gamma_{e-}}{\Gamma_{e-}-1} \beta^2 \gamma^2  \right] - \frac{w_2}{w_1} \nonumber \\
 &&  \qquad- \frac{w_3}{w_4} \frac{\Gamma_{e-}}{\Gamma_{e-}-1} \frac{ \beta \gamma_{12}}{ |\beta_{12} | + \beta} \left[ ( |\beta_{12}  |+ \beta)^2 \gamma^2 - \frac{\sigma_1  |\beta_{12} |}{2\beta^2} \left\{  |\beta_{12} | (1+ \beta^2) + 2 \beta \right\} \right] = 0,
\end{eqnarray}
which is an algebraic equation of fifth order in \(\beta\). It is only slightly simplified by the assumption of equal downstream densities, applying \(w_2 / w_1 = 1\). For a highly relativistic approximation \(\gamma_{12} \gg 1\), Equation (\ref{app:exact}) reduces to a quadratic equation in the shock speed
\begin{equation}
	\beta^2 (1+\sigma_1) \left(1- \frac{w_3}{w_4} \frac{\Gamma_{e-}}{\Gamma_{e-}-1} \right) + \beta \left[ 1+ \frac{\sigma_1}{2} \frac{w_3}{w_4} \frac{\Gamma_{e-}}{\Gamma_{e-}-1} - \frac{1}{\gamma_{12} } \frac{w_2}{w_1} \right] - \sigma_1 \left(1 - \frac{w_3}{2w_4} \frac{\Gamma_{e-}}{\Gamma_{e-}-1} \right) = 0.
\end{equation}

\section{Tables}

\begin{figure}[ht!]
\begin{center}
\includegraphics{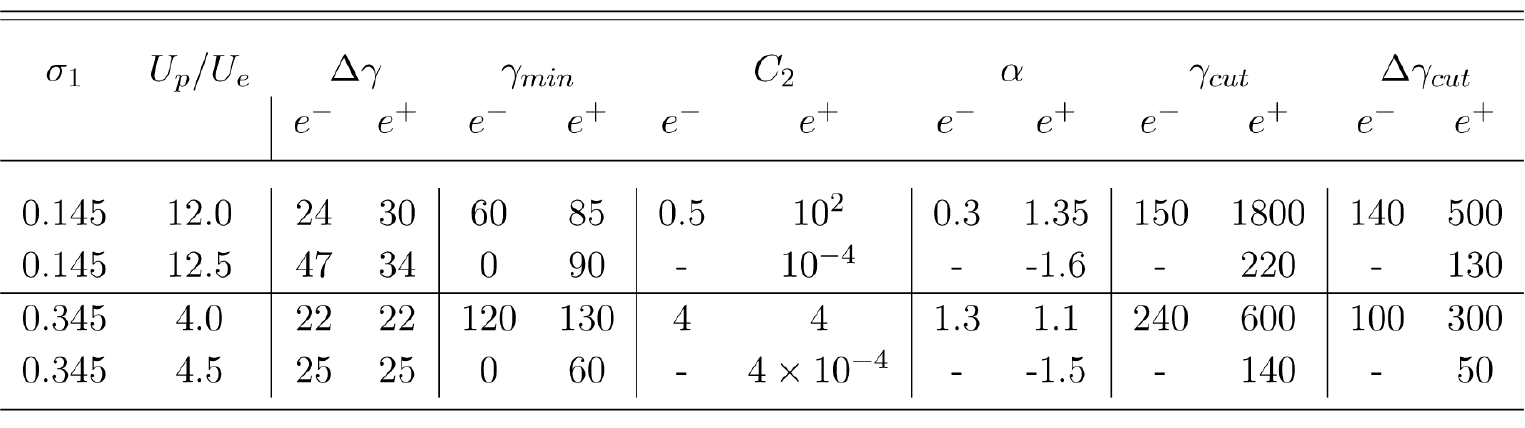}
\end{center}
\vspace{-12pt}
\caption{Downstream parameters measured from simulation data and comparison of density compression and shock speed with theory in brackets, obtained from Equations (\ref{exact}) and (\ref{densityjump}).}\label{tabrange}
\end{figure}

\end{document}